\documentclass[a4paper,fleqn,usenatbib]{mnras}

\usepackage{newtxtext,newtxmath}

\usepackage[T1]{fontenc}
\usepackage{ae,aecompl}

\usepackage{graphicx}	
\usepackage{amsmath}	
\usepackage{amssymb}	

\title[Orbital Period Changes of DQ Her and BT Mon]{Sudden and Steady Orbital Period Changes Across the Classical Nova Eruptions of DQ Her and BT Mon}

\author[B. E. Schaefer et al.]{Bradley E. Schaefer$^{1}$\thanks{E-mail: schaefer@lsu.edu}
\\
$^{1}$Department of Physics and Astronomy, Louisiana State University, Baton Rouge, Louisiana, 70820, USA\\
}

\date{Accepted XXX. Received YYY; in original form ZZZ}

\pubyear{2019}

\begin{document}
\label{firstpage}
\pagerange{\pageref{firstpage}--\pageref{lastpage}}
\maketitle

\begin{abstract}
I report two new measures of the sudden change in the orbital period ($P$) across the nova eruption ($\Delta$P) and the steady period change in quiescence ($\dot{P}$) for classical novae (CNe) DQ Her and BT Mon.  The fractional changes ($\Delta$P/P) in parts-per-million (ppm) are $-$4.46$\pm$0.03 for DQ Her and +39.6$\pm$0.5 for BT Mon.  For BT Mon, the $\Delta$P/P value is not large enough (i.e., $>$1580 ppm) to allow for Hibernation in this system.  The {\it negative} $\Delta$P/P for DQ Her is a confident counterexample of the Hibernation model for the evolution of cataclysmic variables.  Further, published models of period changes by nova eruptions do not allow for such a negative value, so some additional mechanism is required, with this perhaps being due to asymmetric ejection of material.  My program has also measured the first long-term $\dot{P}$ for CNe, with 0.00$\pm$0.02 for DQ Her and $-$2.3$\pm$0.1 for BT Mon, all with units of $10^{-11}$ days/cycle.  These can be directly compared to the predictions of the Magnetic Braking model, where the long-term average $\dot{P}$ is a single universal function of $P$.  The predicted values are -0.027 for DQ Her and -0.33 for BT Mon.  For both novae, the measured $\dot{P}$ is significantly far from the predictions for Magnetic Braking.  Further, the observed $\Delta$P for BT Mon imposes an additional {\it positive} period change of +0.60$\times$10$^{-11}$ days/cycle when averaged over the eruption cycle, so this system actually has a long-term {\it rise} in $P$.
\end{abstract}




\section{Introduction}

Classical novae (CNe) are the most prominent class of cataclysmic variables (CVs), wherein a close binary system consists of a nearly-normal companion star that fills its Roche lobe and spills matter onto an accretion disk then onto a white dwarf (WD).  CNe suffer enormous eruptions when the material accumulated onto the surface of the WD reaches a critical mass and ignites in a thermonuclear explosion, with the system brightening by 8--20 magnitudes and ejecting a mass ($M_{ejecta}$) that is highly uncertain and might be anywhere from 10$^{-4}$ to 10$^{-8}$ $M_{\odot}$.  CNe suffer these nova eruptions with recurrence time scales ($\tau_{rec}$) from one century to around a million years, although recurrent novae (RNe) are an extreme class of novae with $\tau_{rec}$$<$100 years.

The orbital period ($P$) of a nova system is its single most important measureable property.  This largely determines the nature of the system, and it determines the accretion rate and the size of the companion.  All modeling of CNe requires $P$ as the basis for all calculations.  Fortunately, $P$ is usually easy to measure with high accuracy, either from a radial velocity curve or from the periodicity of the photometric modulation.  Except for systems viewed near face-on, all CNe show brightness changes tied to the orbital period, from eclipses or the beaming pattern of the disk light (especially from the hot spot where the accretion stream hits the disk), as well as from ellipsoidal and irradiation effects.  The importance of $P$ for CNe has motivated a vast effort by hundreds of workers (from the 1960s and on-going) to measure and remeasure the photometric modulations on the orbital periods of most CNe.  

{\it Changes} in the orbital period are important because they drive the evolution of CNe, and because they are the key markers of various physical processes in the systems.  These changes can be slow and steady changes throughout the quiescence between eruptions.  For example, the steady period changes ($\dot{P}$ in units of days per cycle resulting in a parabolic $O-C$ curve) that arise from the losses of angular momentum in the binary orbit associated with magnetic braking are thought to drive the long-term evolution of CNe and all CVs (Rappaport, Joss \& Webbink 1982, Patterson 1984, Knigge, Baraffe \& Patterson 2011).  The period changes can also be sudden, at the time of the nova eruption, making for the system going from a pre-eruption orbital period ($P_{pre}$) to a post-eruption orbital period ($P_{post}$).  Such period changes by $\Delta$P (i.e., $P_{post}$-$P_{pre}$) certainly arise from the simple loss of $M_{ejecta}$ from the binary.  The short-term and middle-term evolution of CNe and CVs has long been taken to be driven by $\Delta$P$>$0, with the period and binary separation {\it increasing} across the nova event, leading to a large drop in the accretion rate, as described by the `Hibernation Model' (Shara et al. 1986; Prialnik \& Shara 1986; Livio \& Shara 1987; Shara 1989).

A lot is riding on the $\Delta$P values for novae.  Unfortunately, it is hard to get $P_{pre}$, because the pre-nova system has never been studied.  (No one knew that the anonymous faint star would later go nova.)  The only dodge for CNe is to use archival data that unknowingly measured the star's brightness from {\it before} the nova event, and spot the periodic modulation.  But existing archival data are adequate to pull out $P_{pre}$ for few systems.  Further, useful archival data sources are isolated, now obscure, and little-used, while it requires specialized knowledge and experience to pull out $P_{pre}$ correctly.  Few living astronomers retain the skills and expertise needed, so I have had no competition.  

Previously, Schaefer \& Patterson (1983, SP83) reported one $\Delta$P value for BT Mon and no value for DQ Her.  My more recent publications give the result for V1017 Sgr (Salazar et al. 2017) and QZ Aur (Schaefer et al. 2019).  Further, in a companion paper, I report on measures of two more CNe, RR Pic and HR Del.  This brings to a total of six CNe with measured $\Delta$P, and this is the entire set of possible measures for a decade or more to come.  These are part of my extensive program to measure $\Delta$P for all possible CNe and for many RN eruptions.

This paper starts with a summary of the theory behind the many mechanisms that make for period changes in CNe (Section 2).  A summary of my entire $\Delta$P program is in Section 3.  Section 4 gives a detailed description of the photometry of archival plates, as few astronomers now have any idea of what is going on here, while this is critical for the $\Delta$P program.  The next two sections give my new results for $\Delta$P and $\dot{P}$ for DQ Her and BT Mon.  For DQ Her, a variety of advances (chiefly adding critical new plate material) allows me to finally measure $P_{pre}$ (Section 5).  For BT Mon, I have tested and confirmed the SP83 $\Delta$P with an order of magnitude improvement in accuracy, plus I have added the $\dot{P}$ (Section 6).  In the final section, I use my new measures to address the predictions made by the venerable models of Hibernation and Magnetic Braking.

\section{Period Change Mechanisms For Novae}

Many period change mechanisms are operating for novae, I count six that work during the time of quiescence and four that work from the eruption.  The purpose of my $\Delta$P program is to get unique measures of the period changes, so I have to be careful to recognize what I derive in the $O-C$ diagrams.  For example, for both DQ Her and BT Mon, the steady $\dot{P}$ must be measured so that the post-eruption eclipse times and epochs can be extrapolated back to the time of the nova.  Further, the values of $\dot{P}$ are interesting by themselves, and we will have to recognize the mechanisms that average out to zero in the long term so that these are not confused for evolutionary effects. 

The ephemeris of the system will model the time for some particular orbital phase (like an eclipse or the time of maximum brightness) as 
\begin{equation}
T = E + PN + 0.5\dot{P}N^2,
\end{equation}
where $N$ is an integer counting cycles from the time of the fiducial epoch $E$ (in heliocentric Julian days or HJD).  The epoch selected for being close to the time of the nova eruption will be labelled $E_0$.  With this, the units of $\dot{P}$ are in days per cycle, equalling the period change (in days) over each orbital cycle.  For various physical equations involving $\dot{P}$, as below, the natural units are seconds/second (or days/day or dimensionless), Where the values with the two separate units are related by a factor of $P$, so the readers must be careful to compare $\dot{P}$ values in the same units.  For a sudden period change across the nova event, two applications of this equation are needed, with the pre-nova equation using $P_{pre}$ and $E_0$, the post-nova equation using $P_{post}$ and necessarily the same $E_0$, and both presumably using the same $\dot{P}$.

\subsection{Period Change In Quiescence}

The orbital period of CNe and CVs in quiescence (i.e., far from any nova event) can change for a variety of known physical mechanisms.  On average and over the long term, CV systems must be evolving from long-to-short $P$.

For systems above the `period gap' (roughly 2$<$$P$$<$3 hours), the generally-regarded dominant mechanism is termed `magnetic braking', and this always leads to the slow grinding down of the orbit.  The magnetic braking mechanism starts with an ordinary stellar wind ejected by the companion star.  The companion will have some sort of a magnetic field, the outgoing particles will be tied to the field, and the field will be tied to the rotation of the companion, so out to some distance from the companion, the particles will be forced to start rotating around the companion.  These rotating particles will carry away angular momentum from the rotation of the companion.  The rotation of the companion is tied to the orbital period as it must be synchronously rotating, so the loss of angular momentum by the companion star speedily becomes a loss in the orbital angular momentum of the system.  With this, $P$ must decrease, and the two stars must draw together in their orbit.  This steady robbing of the angular momentum makes for a slow inevitable grinding down of $P$.  At any given time, this should appear as a nearly-constant $\dot{P}_{mb}$ that is negative, with a parabolic $O-C$ curve that is concave down.

A similar mechanism arises from gravitational radiation within the model of General Relativity (GR), as the stars spin around each other, with the emitted gravitational radiation carrying away orbital angular momentum, inexorably grinding down the orbit.  Gravitational radiation is a very weak mechanism compared to magnetic braking, and so is negligible for systems above the period gap.  Below the period gap, the companion star's magnetic activity nearly ceases, so the gravitational radiation becomes dominant.  The negative $\dot{P}$ is precisely prescribed for given stellar masses and $P$.  

Another quiescent-$\dot{P}$ mechanism arises from the simple transfer of mass between the stars as part of the accretion process.  For conservative mass transfer where the system as a whole does not lose mass, the steady period change will be 
\begin{equation}
\dot{P}_{mt}=(3P\dot{M}/M_{comp})(1-q).  
\end{equation}
Here, the companion star has mass $M_{comp}$, the mass transfer rate ($\dot{M}$) is a positive quantity, and $q$ is the usual mass ratio ($M_{comp}/M_{WD}$).  To get $\dot{P}_{mt}$ in units of days/cycle, $\dot{M}$ must be converted from its usual units of $M_{\odot}$/year to units of $M_{\odot}$/cycle.  Non-conservative mass transfer can change this somewhat, but not greatly for practical situations.  For virtually all CNe, $q<1$, so $\dot{P}_{mt}$$>$0 and the orbit is steady increasing in period.  This effect is additive with the effect of magnetic braking.

The above mechanisms all operate over very long time scales, with small and effectively-constant $\dot{P}$.  But real systems are occasionally seen to have very small period changes with time scales of years.  These are both positive and negative.  The observational base for these fast effects are poorly known, mainly because it requires a large number of well-spaced timings over long intervals, all with very high accuracy, so as to see the the effects.

The so-called `Applegate mechanism' (Applegate \& Patterson 1987) connects changes in the  quadrupole moment of the companion's magnetic field (due to a stellar activity cycle) to the orbital period.  The prediction is that this mechanism can produce small changes in $P$ with a periodicity of order a decade or so. I do not know of any confirmed case where the Applegate mechanism is detected, but this is a hard task because the starspot cycles are on time scales of a decade or more, and many cycles of very precise timings are required.  These changes are always small, and must average out to zero in the long-term.

Let me propose a simple mechanism that uses well-known effects that should produce small (apparent) period changes that are both fast and slow, yet average out to zero.  The mechanism is just the ordinary shifting of the `beaming pattern' of the disk light to different directions as the hot spot at the edge of the disk moves around in its usual way.  So, if the hot spot shifts by, say, 10$\degr$ around the circumference of the accretion disk (compared to the direction of the companion star), then the time of maximum brightness will shift by 2.8\% (i.e., 10$\degr$/360$\degr$) of the orbital period in the $O-C$ diagram.  Hot spots are all the time moving back and forth around the edge of the disk (as the accretion rate varies), so this effect will always be present.  It is not that the orbital period (say, from conjunction to conjunction) is changing, but rather that the time-of-maximum-brightness has a variable offset to the time of conjunction.  (This mechanism will not change eclipse times.)  Only in the very best observational cases can this effect be proven, although this will always provide a small noise in an $O-C$ curve where the times are based on the photometric modulation.

Another possible source of apparent period changes comes from third body effects.  For example, as the binary moves towards-and-away from the Earth in its orbit with a tertiary star, the light travel time effects will make for a sinusoidal variation in the $O-C$ curve, superposed on the real period changes.  When viewed over a fraction of the third body's orbital period, the light travel time effects will look like a parabola in the $O-C$ curve.  I know of no confirmed case of third body period changes in any CN.

For the purposes of this paper, the short-term mechanisms all average out to zero in the long term, so they can be ignored for consideration of CV evolution.  Rather, they will appear as noise superposed on the parabolic $O-C$ curve arising from the long-term mechanisms.  The gravitational radiation mechanism is negligible for the long-period CNe in this paper.  So we are expecting a parabolic $O-C$ curve over the decades, with some noise superposed, with the $\dot{P}$ being a balance between the positive mass transfer effect and the negative magnetic braking effect.

Century-long measures of the observed $\dot{P}$ (i.e., the total steady change for evolution) are a needed by-product of my $\Delta$P program.  I know of no prior measures of the real long-term $\dot{P}$.  Well, I have $\dot{P}$ between eruptions for {\it recurrent} novae U Sco, CI Aql, T CrB, and T Pyx (e.g., Schaefer 2011), including that the parabolic term in the $O-C$ curve changing by one order-of-magnitude from before the 2010 eruption of U Sco to after.  For classical novae, Pringle (1975) and Beuermann \& Pakull (1984) constructed an O-C diagram for T Aur from 1962--1982, but the intrinsic scatter was much too large for their short interval, so they were just fitting a parabola to noise.  Presumably, someone could update their list of eclipse timings to 2019 and maybe obtain a real measure of the evolutionary $\dot{P}$, but no one has done so.  Vogt et al. (2017) has constructed an $O-C$ curve for RR Pic, fitting lines, parabolas, and sinewaves, but none of these show the real evolutionary effect, which is only seen when the $O-C$ curve is greatly extended in time (see the companion paper).  Various workers constructed $O-C$ diagrams for DQ Her (Patterson, Robinson, \& Nather 1978, Africano \& Olson 1981, Zhang et al. 1995), but their modest time intervals were relatively short and they were mainly looking at the fast variations, missing any parabolic term.  So, my $\Delta$P program is producing the first measures of the long-term evolutionary component of $\dot{P}$ for classical novae.

\subsection{Period Change Across The Eruption}

The nova orbital period can also change suddenly at the time of the nova eruption, from $P_{pre}$ to $P_{post}$.  The original mechanism (Ahnert 1960) was simply the mass loss from the nova system, where Kepler's Law requires that the orbital period must get longer after an eruption.  With detailed balancing of angular momentum, many workers have derived the $\Delta$P from mass loss as
\begin{equation}
\Delta P_{ml} = A P M_{ejecta} / M_{WD},
\end{equation}
\begin{equation}
A = (2q+3q^2-3q^2\alpha + 3q^2\alpha \beta -3\beta -2\beta q)/(q+q^2),
\end{equation}
\begin{equation}
\beta = - \dot{M}_{comp} / \dot{M}_{WD} = (\pi [R_{comp}/a]^2)/(4 \pi).
\end{equation}
The $\alpha$ parameter gives the average specific angular momentum of $M_{ejecta}$ in units of the WD's angular momentum, and this should be close to unity.  The $\beta$ parameter gives the fraction of $M_{ejecta}$ that lands on the companion star, and this is small.  For the case with $\alpha$=1 and $\beta$=0, then $A=2/(1+q)$, or 
\begin{equation}
\Delta P_{ml}/P = 2 M_{ejecta} / (M_{WD}+M_{comp}),
\end{equation}
For typical stellar masses and for the range of ejecta masses, the fractional change in $P$ might range from 0.1 ppm for RNe up to 100 ppm or so for the most massive nova shells.  With this mechanism, the orbital period must get longer, and the nova event must drive the two stars apart.

During the nova eruption, the outgoing $M_{ejecta}$ creates a shell of gas, which the companion star must plow through as it moves around its orbit.  The `friction' and turbulence in the nova shell will provide a force acting on the companion, slowing it down in its orbital motion.  This takes angular momentum out of the orbit (to be carried away by the outgoing gas), causing a slight in-spiral of the orbit, and suddenly decreasing the orbital period over a time intervals of weeks.  This mechanism is labeled `frictional angular momentum loss' (FAML).  Livio (1991) gives
\begin{equation}
\Delta P_{FAML}/P = -0.75 \frac{M_{ejecta}}{M_{comp}}\frac{V_{comp}}{V_{ejecta}}(R_{comp}/a)^2.
\end{equation}
Given that the orbital velocity of the companion star ($V_{comp}$) is always much less than the expansion velocity of the ejecta ($V_{ejecta}$) for the nova considered here, and that the radius of the companion star ($R_{comp}$) is always substantially less than the semimajor axis of the orbit ($a$), the effects of FAML will always be greatly less than the effects of mass loss.  FAML is driving the system to shorter periods, while mass loss is driving the system to longer periods, but the effects of mass loss always dominate the sum.

Martin, Livio, and Schaefer (2011) propose a third method of getting a sudden period change across the nova event.  They realized that the magnetic field of the companion star will try to force co-rotation of the gas in the expanding shell, and that this will carry away some of the rotational angular momentum of the companion.  The companion star's rotation is speedily synchronized with the orbit, so the net result is that the orbit loses angular momentum, and the period suddenly decreases.  (This is much the same as the magnetic braking mechanism during quiescence, except that the gas involved is from the nova shell rather than the companion's wind.)  I will label the period change from this effect as $\Delta$P$_{mbns}$, with the subscript standing for `magnetic breaking in the nova shell'.  For $q$$<$1, the orbital period will decrease due to this effect.  For the mass ratios relevant for most CNe, the total period change ($\Delta$P$_{ml}$+$\Delta$P$_{FAML}$+$\Delta$P$_{mbns}$) will be positive unless that companion star has such a high magnetic field that its Alfven radius is larger than three-quarters of the semi-major axis.  This would require an unreasonably high magnetic field for the companion.  Thus, for the practical cases of CNe, the first three mechanisms combined can only makes $\Delta$P/P to be positive.  This is a critical point, as my $\Delta$P program is finding that most CNe have large negative $\Delta$P/P.  

It is important to explore all possible mechanisms to create period changes across nova eruptions.  We can consider the inevitable effects wherein the nova eruption itself irradiates the companion star, whose atmosphere expands slightly to overfill the Roche lobe, thus increasing the mass transfer only during the eruption duration, and this extra mass transfer will drive an additional $\Delta$P.  For the CNe with $q$$<$1, the effects must be to increase the orbital period (i.e., contribute a positive $\Delta$P) over the duration of nova eruption.  So this mechanism cannot account for the large negative $\Delta$P seen in 5-out-of-6 of the measured CNe.  The size of the effect can be estimated from the calculations of Kovetz, Prialnik, \& Shara (1988), where they calculate the extra mass transfer coming from the irradiation of the low-mass companion stars in novae systems.  For a 0.75 M$_{\odot}$ companion, the accretion rate in a typical nova event is 6$\times$10$^{-9}$ M$_{\odot}$/yr at a time 1.0 years after the start of the eruption, and it varies as the -0.45 power of time.  The overflow mass during the nova's massive wind phase will be entrained and ejected.  The integrated mass transferred from 0.1 year to 10 years is 2$\times$10$^{-8}$ M$_{\odot}$.   For the usual case of mass transfer, this mass makes for a fractional period change of roughly 0.03 ppm.  Thus, this effect is negligibly small in all cases.

So we are needing some mechanism that can greatly shorten the orbital period, with a size substantially larger than possible for $\Delta$P$_{ml}$ and in the opposite direction.  J. Frank has proposed (in Schaefer et al. 2019) just such a mechanism that can easily produce large negative $\Delta$P, and this mechanism is observed to be ubiquitous.  The idea is that nova shells are always asymmetric in images, so the WD is throwing more matter outward in some directions, and such can act as a weak and wide `jet' that can speed-or-slow the orbital motion of the WD.  For a schematic model, assume that the ejecta consists of two hemispherical shells with $M_{forward}$ and $M_{back}$ going in the forward and backwards direction of the instantaneous motion of the WD.  An asymmetry parameter is $\xi$=$(M_{forward}-M_{back})/M_{ejecta}$, which can run anywhere from -1 to +1.  Imaging of nova shells often shows large asymmetries to one side, so that the magnitude of $\xi$ often has lower limits of 0.4.  The period change from this `jetting' effect is
\begin{equation}
\frac{\Delta P_{jet}}{P} = \frac{M_{ejecta}}{(M_{WD}+M_{comp})}\left(- q\xi\frac{3 V_{\rm ejecta}}{ 2V_{WD}}\right)\, .
\end{equation}
For a typical WD orbital velocity (say, $\sim$100 km/s), a typical CN ejection velocity (say, $\sim$1000 km/s), and a typical mass ratio (say, $\sim$0.8), a $\xi$ larger than 0.2 will make for the total $\Delta$P/P to be negative.  With large $\xi$ values (as seen for some nova shells) and a massive nova shell, we can get a total $\Delta$P/P as much as -1000 ppm.  The expected size of this effects depends on a variety of details relating to the mass and velocity imbalances in the ejecta, and the directionality with respect to the WD orbital motion.  This mechanism can make $\Delta$P/P be large and negative (as seen for various of my CNe), or large and positive (if the `jetting' is pointed so as to speed up the WD) or with only small change (as for BT Mon and DQ Her).  We know that these weak-and-wide jetting effects are common and often large for CNe, and the jetting might well dominate over the other three mechanisms.

\subsection{The Hibernation Model}

The Hibernation Model (Shara et al. 1986; Shara 1989) for CV evolution described the changes over a many millennia cycle, as the system varies from a high-accretion nova system down to a low-accretion dwarf nova, and then to a disconnected binary.  This cycle is started and driven by a large positive $\Delta$P, which forces the two stars apart.  After the effects of the nova eruption fade away, the accretion rate fall precipitously to low levels due to the stars being suddenly separated.  While the accretion rate is falling, the system will pass through states as a novae-like star, a Z Cam system, an ordinary dwarf nova, and then a disconnected close-binary with little accretion.  From this disconnected state, magnetic braking will keep grinding down the orbit, making the stars in-spiral until they come into contact and accretion picks up again.  As the accretion rate rises, the star shifts to appearing as a dwarf nova, a Z Cam star, then a nova-like system.  Finally, as a nova-like system, enough material will accumulate on the WD so that another nova eruption happens.  Then the cycle starts all over.

This is a wonderfully evocative model that is physically plausible.  And it has some nice successes; including the ability to explain the very wide range of accretion rates for similar CVs with the same $P$, and accounting for some key discrepancies in the space densities of CVs.  Another successful prediction is that systems now appearing as various types of dwarf novae are seen to have very old CN shells (Shara et al. 2007; 2012; 2017), consistent with the now-low-$\dot{M}$ systems being high-$\dot{M}$ nova systems many centuries ago.  (These nova shells are also easily consistent with the default no-evolution case, where the long-ago dwarf nova finally accumulated enough material on the white dwarf to have a nova eruption, and now, the system still appears as a dwarf nova yet with a nova shell.   So the shells around dwarf novae are not useful evidence for the Hibernation model.)  This Hibernation cycle is superposed on the long-term evolution imposed by magnetic braking.  Starting with its proposal in 1986, with no well-developed alternatives, Hibernation has remained the primary idea for middle-term CV evolution.  

But Hibernation has not convinced our community (e.g., Naylor et al. 1992; Weight et al. 1994; Warner 2008; Thomas, Naylor \& Norton 2008).  The fundamental problem is that Hibernation is predicting changes that happen on time scales of a century or longer, and such cannot be tested with modern telescopes or satellites, because they only take snapshots in time with any changes from Hibernation unobservable.  (This is why archival data are the critical path for testing Hibernation, because only archival data can test predictions of long-term changes.)  Variations on Hibernation have been proposed, with differing order of CV classes and with the drop in accretion not being so deep (Livio, Shankar, \& Truran 1988; Vogt 1989; Livio 1989; Patterson et al. 2013).

A critical prediction and requirement and driving force for Hibernation is that $\Delta$P$>$0, so that the binary will necessarily separate and the accretion must then drop.  For most of the last three decades, we only had the case of BT Mon, which satisfies the minimal Hibernation prediction.  Then Salazar et al. (2017) and Schaefer et al. (2019) found two CNe for which the $\Delta$P is large and negative, with both observational results being robust and confident.  This provides a certain proof that Hibernation is not operating for these two systems.  However, before we can fully reject the now-venerable model of Hibernation, we should have more tests of the basic prediction.

But the requirement for Hibernation to be operating is actually greatly more restrictive, as $\Delta$P must really be quite large so as to get a substantial drop in accretion.  For the ideal Hibernation result that the binary disconnects, the system really has to drop its absolute magnitude in quiescence from $M_{V,q}$ fading to a hibernation value of $M_{V,hib}$=+12 or fainter.  (This is the lower limit of recognized CVs, as taken from the {\it Gaia} DR2 data.)  Or if the Hibernation does not drive the system to disconnection (in some sort of a shallow hibernation model), a fading of $M_{V,q}$ by less than 5 mag or so would not be considered as something we would call 'hibernation'.  So we have two criteria, where the absolute magnitude in the hibernation state ($M_{V,hib}$) either is fainter than +12 mag or where $M_{V,hib}$-$M_{V,q}$$\geq$5.

These criteria for Hibernation can be directly related to $\Delta$P.  With the accretion light dominating the system brightness, which is roughly proportional to the accretion rate, \begin{equation}
M_{V,hib}-M_{V,q}=~2.5 \times \log_{10}(\dot{M}/\dot{M}_{hib}).
\end{equation}
The ratio of the accretion rate in the current state to the ultimate hibernating state is 
\begin{equation}
\dot{M}/\dot{M}_{hib}=e^{\Delta R_{comp}/H}, 
\end{equation}
where $H$ is the atmospheric scale height (Eq. 4.19 in Frank et al. 2002).  For this derivation, $M_{V,hib}$ is taken to be the luminosity from the accretion only, and $\dot{M}_{hib}$ must always be positive.  The change in the effective radius of the Roche surface is \begin{equation}
\frac{\Delta R_{comp}}{R_{comp}}=\frac{2}{3}\frac{\Delta P}{P}+\frac{1}{3} \frac{M_{ejecta}}{M_{comp}}.
\end{equation}
We now have an accurate and fairly simple relation between the observed period change and the drop in brightness:
\begin{equation}
M_{V,hib}-M_{V,q} = 1.086 \frac{R_{comp}}{H}[(\frac{2}{3} \frac{\Delta P}{P})+(\frac{1}{3} \frac{M_{ejecta}}{M_{comp}})].
\end{equation}
With this, I can calculate the minimum $\Delta$P/P to satisfy the two criteria.  

To illustrate this, let us examine the case of what is needed to allow shallow hibernation.  This requires $M_{V,hib}$-$M_{V,q}$$\geq$5.  For a CN with $M_{ejecta}$ near the largest value of the expected range, we would get a maximum value something like $\frac{M_{ejecta}}{M_{comp}}$ of 10$^{-5}$.  Typical values of $\frac{R_{comp}}{H}$ are near 5000, with this being applicable to BT Mon, DQ Her, and V1017 Sgr.  With this, to get minimal hibernation, we see that $\Delta$P/P must be greater than +1380 ppm.  This is large, very large.  From the above mechanisms, this is impossibly large.  So even minimal Hibernation is not possible.

\subsection{The Magnetic Braking Model}

The Magnetic Braking Model (MBM) is now very well developed, and is widely accepted as the general description of the long-term evolution of CVs.  The model makes a lot of sense, and it has enjoyed various successes.  For example, it explains the general shift in $\dot{M}$ as a function of $P$, it explains the period gap, and it explains the minimum $P$ seen for CVs below the period gap.  The on-going discussions are now about things like tweaking the efficiency of the the braking, and wondering whether there is any residual magnetic braking below the period gap (e.g., Knigge et al. 2011; Schreiber, Zorotovic \& Wijnen 2016).  The model has been around for many decades, and it can be described as being `venerable'.  

Importantly, the strong success of the MBM only rely on the general scenario of `angular momentum loss' (AML) of the binary as it grinds down the orbit.  The values and functions for the magnetic braking mechanism in particular are poorly known, with published proposals varying by over two orders-of-magnitude for CVs above the period gap (Knigge et al. 2011, Fig. 2).  Further, there is neither a knowledge nor a need for the AML to be dominated by the particular magnetic braking mechanism.  To achieve the great successes, all we need is that the sum of all the AML mechanisms average out to some particular level.

The current standard MBM calculate the $\dot{P}$ from 
\begin{equation}
\frac{\dot{P}_{model}}{P}= 3 \frac{\dot{J}}{J} +3 \frac{\dot{M}}{M_{comp}}(1-q),
\end{equation}
for conservative mass transfer.  Again, the $\dot{M}$ is taken to be positive, as the mass accretion rate.  Again, the natural units for $\dot{P}$ in this equation are days/day (or dimensionless), so this must be converted to units of days/cycle for comparison with my observed values.  Here, $J$ is the total angular momentum of the system.  The rate of change of $J$ is $\dot{J}$, which will be negative for the AML mechanisms.  The $\dot{J}$ will come from two mechanisms, magnetic braking and GR.  For the CNe in this paper, with $P$ larger than the upper edge of the period gap, the General Relativistic effects are negligibly small.  The two terms in this equation can be usefully labeled as coming from the magnetic breaking mechanism and from the effects of the accretion mass transfer.  With this, the model period change can be seen as $\dot{P}_{model}=\dot{P}_{mb}+\dot{P}_{mt}$.  The total model period change is a balance between the negative $\dot{P}_{mb}$ and the positive $\dot{P}_{mt}$, both of which have comparable magnitudes.

Knigge et al. (2011) have calculated the required level of $\dot{J}$ so as to match the observed edges of the period gap (2.15--3.18 hours) and the bounce period (82.4 minutes).  They adopt a fiducial model as the AML prescription from the $\gamma$=3 model in Rappaport, Verbunt \& Joss (1983), which they label as the `standard' model.  They calculate that the observed period gap and bounce period is matched for an AML mechanism that has $\dot{J}$ being a factor of 0.66 times their standard model, and the total AML below the period gap that requires a factor of 2.47 times the $\dot{J}$ from GR.  With the GR contribution being confidently known, this means that there is some additional AML mechanism operating below the period gap that acts with a strength around 1.47 times the GR effect.  A reasonable idea for this is that the secondary star might retain some residual magnetic braking even below the period gap.  Above the period gap, with the effects of the magnetic braking mechanism known only to order-of-magnitude, the Knigge et al. (2011) calculations are only showing us the level at which AML must be operating on average, from any and all AML mechanisms.

The MBM requires that the long-term average $P$ be steadily decreasing, with $\dot{P}$$<$0 as prescribed by the detailed physics (Rappaport et al. 1982, Patterson 1984, Knigge et al. 2011).  This prediction is testable, but only if we can get a long run of precise timing data.  Confounding this test are the various short-term effects that make for apparent noise in the $O-C$ diagram around some parabolic curve.  At this time I do not know of any CN cases where the long-term $\dot{P}$ has been significantly measured above the noise.  In my $\Delta$P program, I am getting long-term $\dot{P}$ values as a needed by-product, so my $O-C$ curves can also be used to measure and test the claims of the MBM.

A further issue that arises for the MBM comes from the $\Delta$P for each individual eruption.  The period change across each nova event contributes to the long-term change also, with an effective $\dot{P}_{\Delta P}$ that is averaged over the entire eruption cycle, so 
\begin{equation}
\dot{P}_{\Delta P} =  \Delta P/(\tau_{rec}/P),
\end{equation}
with units of days per cycle.  Such effects are not included in any current MBM models, simply because the realization is only recent that these effects are large and outside the ordinary effects of mass loss.  If $\dot{P}_{\Delta P}$  is large and negative, then this mechanism will dominate over the $\dot{P}$ from magnetic braking ($\dot{P_{mb}}$), and the effects of magnetic braking will become negligible.  If $\dot{P}_{\Delta P}$  is large and positive (as required by Hibernation), then the evolution would not even be driven towards shorter periods.  In either case, the MBM must give greatly wrong answers.  So it is important to measure typical values of $\dot{P}_{\Delta P}$  for comparison with $\dot{P}_{mb}$ and $\dot{P}_{mt}$.

The MBM requires that all CVs follow a single particular evolutionary path, that can be described as `universal'.  This path can be exemplified as the curve of $\dot{M}$ versus $P$, with the inevitable grind-down of the orbit and the accretion rate, with a break in accretion around the period gap, and a bounce in period where the very old CVs start increasing $P$.  This picture is known to be too simplistic, with the primary proof being that CVs with a given period actually have a very wide range of $\dot{M}$.  To get around this contradiction, various short-term and middle-term evolutionary cycles are invoked (Hibernation is the most prominent model for this), with these being superposed on the very-long-term trend described by the MBM.  That is, the model is only describing evolution on time scales of millions and billions of years, while real CVs vary about this ideal for various poorly understood reasons.  Thus, CNe and CVs only follow the MBM when the measured values are averaged over a suitably long interval.

The MBM makes very specific predictions about the `universal' path for how the orbital period changes over time, so we have a particular prediction of $\dot{P}$ for any given $P$.  For the AML levels required to match the period gap and the bounce period, Knigge et al. (2011) have calculated $\dot{P}_{model}$ as a function of $P$, as displayed in their Fig. 11 and 12.  (Again, their quoted $\dot{P}$ values are dimensionless and must be converted to units of days/cycle for comparison with my observations.)  These are the MBM predictions that I can test with my measures of $\Delta$P and $\dot{P}$.

\section{The $\Delta$P Program}

Since 1983, I have been working on a long-term program to measure the changes in the orbital periods across nova eruptions.  The basic work is to obtain some `modern' measure of the orbital period after the nova ($P_{post}$), and then to measure the period before the eruption ($P_{pre}$) by finding the periodic photometric modulation in archival data.  J. Patterson first told me of the basic idea when he asked me to use the collection of archival sky photographs (plates) at Harvard to find the orbital period change of BT Mon (SP83).  Ahnert (1960) had previously used the same method to claim that he had measured $\Delta$P for the 1934 eruption of DQ Her.  SP83 also searched for $P_{pre}$ for DQ Her, showed that Ahnert's proposed value was not correct, but could find no alternative.

After this work for two CNe, I began working with RNe.  My idea was that RNe are the only nova systems for which we confidently knew that a particular star will sooner-or-later go nova, so we now have time to measure $P_{pre}$, await the upcoming eruption, then spend another long time measuring $P_{post}$.  So I started seeking periods of RNe, and making many long time series light curves to define $P$ to high accuracy.  To get high accuracy in the period, I need many years of light curves.  This was knowingly a program requiring several decades of work.  This effort has used roughly 300 nights of telescope time, and has also used archival magnitudes taken on 12 trips to 7 observatories.    Sample papers are Schaefer (1990; 2009; 2011) and Schaefer et al. (1992; 2013).  This RN part of my $\Delta$P program is still on-going.

The original motivation for measuring $\Delta$P for RNe was to derive $M_{ejecta}$ for each eruption.  Given Kepler's Law for an idealized system, $M_{ejecta}$ will be proportional to $M_{WD}$($\Delta$P/P), with the constant of proportionality known tightly and near to unity.  $M_{ejecta}$ could then be compared to the mass accreted between eruptions ($M_{accreted}$) to see whether the WD was gaining or losing mass.  ($M_{accreted}$ could be reasonably estimated as the product of the recurrence time scale, $\tau_{rec}$, and the accretion rate, $\dot{M}$.)  There is a balancing act between the mass lost and gained by the WD, with this balance being unknown.  If the near-Chandrasekhar-mass WD is gaining mass over each eruption cycle (i.e.., $M_{ejecta}$<$M_{accreted}$), then it inevitably must reach the Chandrasekhar mass, collapse, and apparently suffer a thermonuclear explosion as a Type Ia supernova.  Thus, for many years from the 1970s up until a few years ago, RNe were regarded as the primary single-degenerate path for solving the important Type Ia progenitor problem.  A measure of $\Delta$P for RNe thus became a critical test (and disproof) for the most prominent solution to one of the most important questions in all stellar astronomy.

RNe are a particularly important subset of novae.  While they might share the same physical mechanisms and settings as CNe, the critical system parameters are all greatly different (Schaefer 2010).  These differences include (1) RNe have near-Chandrasekhar-mass WDs versus CNe usually have much lower mass WDs, (2) most RNe have huge evolved companion stars versus the CNe that almost all have small, low-mass, near-main-sequence companions, (3) most RNe have their accretion driven by the evolutionary expansion of the companion versus CNe where the accretion is driven by angular momentum loss of the orbit, (4) RNe must have very high accretion rates versus CNe that usually have $\sim$100$\times$ smaller rates, (5) RNe ejection velocities are $>$2000 km/s versus CNe with have greatly lower ejection velocities, and (6) RNe have $\tau_{rec}$$<$100 years versus CNe which apparently have longer time scales, usually $\sim$100$\times$ longer.  With all the critical properties of RNe being greatly different from those for CNe, any results, conclusions, or lessons from RNe have only dubious application to CNe.

For CNe, the original physical motivation to measure $\Delta$P was to then deduce the $M_{ejecta}$ from a nova event.  The values of $\Delta$P, $M_{WD}$, and $P$ can all be measured to good or high accuracy, so this promises an accurate measure of $M_{ejecta}$.  Further, the critical measurement is a timing issue and the physics is simple dynamics, so this measure is independent of distance, extinction, filling factors, and much more.  Such high accuracy and confidence is in stark contrast to all previous methods for measuring $M_{ejecta}$ having real uncertainties $>$2 orders-of-magnitude (see Appendix A of Schaefer 2011).  Similarly, theoretical models predicting $M_{ejecta}$ have $>$2 orders-of-magnitude in real uncertainty.  So there is high value in measuring $\Delta$P for CNe.  Further, the evolution of CNe depends on the $\Delta$P across each eruption.  

So starting over a decade ago, I extended my $\Delta$P program from RNe to include CNe, although this is actually just an extension of the original work from 1983 where I sought to measure the $\Delta$P for two CNe (BT Mon and DQ Her, SP83).  The aim is to measure as many $\Delta$P values as possible for CNe.  This can be used as a direct test of the Hibernation Model for CV evolution.

The primary problem for the CN part of my $\Delta$P program is to find a classical nova for which adequate photometric light curves can be obtained before the eruption.  The date of the nova eruption cannot be too early (say, before 1920 or so) or there will be too few pre-eruption plates over too short of an interval, and it cannot be too late (roughly after the year 2000) or else the light curve will not have faded to quiescence for long enough a time to get a good $P_{post}$.  The quiescent brightness must be brighter than roughly 16th mag, or else there will be too few plates to recover $P_{pre}$.  The quiescent nova must display photometric modulations large enough to produce a detectable signal, with the useable amplitudes being $\gtrsim$0.1 mag for a target with many plates and $\gtrsim$0.4 mag for a target with few plates, with this being largely a function of the inclination.  

Let me give examples of novae that cannot be used:  Among the all-time brightest CNe; V603 Aql is not useable because its inclination is 13$\degr$$\pm$2$\degr$ so orbital modulations are overwhelmed by superhumps, GK Per went off in 1901 and has too few pre-eruption plates to be useful, and CP Pup is always too faint to be detected (even to B$>$19.5 on one Harvard plate).  Amongst CNe that eclipse, many have too faint a quiescent counterpart for any useful archival material to be found (DO Aql, V1494 Aql, RR Cha,  BY Cir, DD Cir, CP Cru, V1668 Cyg, V838 Her, V849 Oph, V909 Sgr, and QU Vul), or have too small an amplitude to be detectable (V2540 Oph and V382 Vel), or have too early an eruption to allow for pre-eruption archival material (T Aur, V Per, and WY Sge).

After extensive searching, I have found only six CNe that have adequate pre-eruption light curves for pulling out $P_{pre}$.  V1017 Sgr has a $\Delta$P reported in Salazar et al. (2017), where the period decreased by 273$\pm$61 parts per million (ppm), that is $\Delta$P/P=$-2.73$$\times$10$^{-4}$.  QZ Aur has a $\Delta$P reported in Schaefer et al. (2019), with a period decrease for $\Delta$P/P= $-$290.71$\pm$0.28 ppm.  DQ Her had a false $P_{pre}$ reported by Ahnert (1960), SP83 were unable to pull it out, and I have now made advances to derive a confident pre-eruption orbital period in this paper.  BT Mon has $\Delta$P/P=+39$\pm$4.8 ppm reported in SP83, and I here test, improve, and confirm the original measure.  Further, in a companion paper, I report $\Delta$P and $\dot{P}$ for RR Pic and HR Del.

What about the possibilities of getting more $\Delta$P measures in the future?  The problem is that all currently known novae with adequate pre-eruption data have already had the archival data effectively exhausted, and all currently known novae have already been exhaustively searched for adequate archival data.  So the only possible way to get more $\Delta$P measures is to have new novae erupt, have the infrequent conditions that the currently existing pre-eruption archival data can pull out an accurate $P_{pre}$, then to wait for a decade or more to get an adequate $P_{post}$.  For this future program, the many novae presumably to be discovered by modern surveys will almost all be useless, as their quiescent magnitudes will be too faint (below 17th mag or so) for even modern surveys over the last two decades to pull out $P_{pre}$.  So with typical nova amplitudes, the future successful cases will peak much brighter than roughly 8th mag, and likely be discovered by amateur astronomers.  So starting perhaps a decade from now, we might be able to get another $\Delta$P measure if we get lucky on many points.  But naked eye novae are uncommon, at the 3-per-decade level in recent times, and most of these not having conditions for which $P_{pre}$ can be pulled out, we see that the wait for one or two more $\Delta$P measures is likely to be two or more decades.  Thus, my current $\Delta$P program for CNe has all possible $\Delta$P measures for the foreseeable future.

\section{Photometry with Archival Plates}

The use of archival plates to get $P_{pre}$ is central to my $\Delta$P program.  Unfortunately, the capabilities, properties, and even the existence of the various plate archives worldwide are largely unknown to modern astronomers.  Further, the techniques, analysis, and subtleties of photometry with plates is now largely lost on the world's astronomers.  From the 1890s to the 1980s, many astronomers were expert practitioners, while the majority well-knew the capabilities and techniques.  With the extremely wonderful capabilities of CCD imaging and photometry, the field has correctly made all new research with the electronic tools.  But a cost is that now few living astronomers have the knowledge, capability, or interest to use archival plates, so a whole century of the history of the sky is largely lost.  For many applications, there is no need to know the history of individual targets for more than the decade or so available to modern instrumentation.  But for many applications, such as searching for long-term changes in CN periods, the only way to advance on modern astrophysics questions is to make modern use of the old archival plates.

Roughly three-quarter-of-a-million direct imaging photographic plates are now archived at observatories (mostly in North America and Europe) covering all of the sky.  The premier plate collection is at Harvard, where the entire sky (north and south equally) is covered by half-a-million plates from 1889--1989 (with the infamous Menzel Gap from roughly 1953--1973), with typically 2000 plates recording every star down to 13th mag, hundreds of plates for stars of 15th mag, roughly ten plates on average down to 17th mag, and the deepest plates go below 19th mag.  For coverage before the 1920s, Harvard is the only archive with any coverage.  With this astoundingly good coverage, the Harvard plates are always the primary basis for most any archival program.  The second biggest archive is at Sonneberg Observatory in Germany, with nearly a quarter-of-a-million plates, some coverage of the southern skies, and reaching similar depths as the Harvard plates.  The coverage picks up only in the 1920s, but sky patrols extend even to today.  The Sonneberg plates are a wonderful means to get a very long light curve, even for faint stars.  Other observatories have smaller archives, often filling needed niches, for example the Palomar and ESO Schmidt plates cover the whole sky to 21st mag in two colors, although with few epochs.  (Note, direct visual examination of the plates goes about one mag deeper than the scanned versions.)  Other observatories with good plate archives that I have used for this program are the Maria Mitchell Observatory (with the plates now stored at the {\it Pisgah Astronomical Research Institute} in North Carolina), Asiago Observatory in northern Italy, Bamberg Observatory in Germany, the Vatican Observatory (with the plates now just outside Castel Gandolfo, south of Rome Italy), Jena Observatory in Germany, and the {\it Carte du Ciel} southern plates now at Macquarie University in Sydney Australia.

Almost all of the plate material was recorded on various types of `blue emulsion'.  Before the 1970s, all these types of `blue emulsion' had essentially identical spectral sensitivity.  This means that all blue plates have identical color terms.  The Harvard plates were used to define the photographic magnitude system through the North Polar Sequence and later the Selected Areas.  This set of standards was then used to define the Johnson B system, and then other systems, including the Landolt standards (Landolt 1992; 2009; 2013).  The native system of all the old blue plates is indistinguishable from the Johnson B system.  The old `photographic magnitudes' differ from the Johnson B magnitudes only in that their comparison sequences were distorted.  For stars brighter than 10th or 12th mag, the distortions were small, but the distortions exceed one magnitude for faint stars (Sandage 2001).  Fortunately, if the comparison star sequence magnitudes are in Johnson B, then all the magnitudes taken from the plates will be perfectly a modern B magnitude.  

Before 2010, painstaking individual calibrations from specialized photometry under perfect sky conditions was required to get B-band comparison sequences for stars fainter than 10th mag.  But this all changed with the advent of several systematic all-sky photometry catalogs reaching faint magnitudes.  Now, the {\it American Association of Variable Star Observers} (AAVSO) has an all-sky catalog in BVgri going to 19th mag, with this {\it AAVSO Photometric All-Sky Survey} (APASS) providing the perfect solution for all archival work.  For the comparison sequences of DQ Her and BT Mon in particular, they have already been published is Henden \& Honeycutt (1997), along with finder charts, and the sequences are permanently and publicly available on-line with the AAVSO web site{\footnote{https://www.aavso.org/apps/vsp/}}.  Indeed all the photometry in this paper is now based on APASS comparison sequences.  So when held to this modern standard, the archival magnitudes have a zero color term and zero variations over time.  With the blue emulsions having the Johnson B spectral sensitivity and the APASS B comparison sequences, we can know confidently that the light curves are in the fully modern B system.  

The photometry is always extracted from the plates by a detailed comparison of the effective radius of the target star versus the effective radii of nearby stars (both brighter and dimmer) of known magnitude.  The calibration of the comparison sequence must always be calculated locally for each plate.  The traditional and usually-best way to measure the the relative radii of the stars is visually, by direct comparison when looking through a loupe at the plate on a light table.  For a good plate and a good comparison sequence (the usual conditions), an inexperienced worker will have a one-sigma photometric uncertainty of $\pm$0.3 mag, while an experienced worker will have a one-sigma photometric uncertainty of $\pm$0.10 mag or slightly better.  An example of a typical measurement would be to judge that the target star is, say, one-third of the way from a comparison star of B=12.34 to a comparison star of B=12.67, and deduce a magnitude of B=12.45.  With much practice, this process becomes fast, easy, and accurate.  Various alternative means have been devised to measure the image radii, including `flyspankers', Iris Diaphragm Photometers, and various two-dimensional scanning techniques.  Critically, after many blind experiments, I have always found that none of these alternative methods ever produces an accuracy better than the direct visual examination, and often these alternative methods do worse.  Under some important conditions (the target is close to the plate limiting magnitude, or the target star is crowded with nearby stars), the human eye and experienced judgement is greatly better than any scanning technique.  These alternative methods are always very expensive in time (and sometimes in equipment), and so it is always better to have an experienced eye directly measuring the magnitudes from the plates.

Starting a decade ago, J. Grindlay at Harvard had the vision to realize that the community needs to have all the Harvard plates scanned and publicly available on-line, and he has had the leadership to get his vision working in an excellent manner.  To date, his program ({\it Digital Access to a Sky Century @ Harvard}, or DASCH) has scanned roughly half the plates, now covering half the sky, and performed automated photometry on every star image on all the plates.  So now for half the stars in the sky (down to 17th mag or so), an outside user can quickly download a full B light curve history from 1889 to 1989 (101 years).  Suddenly many modern astrophysics project ideas become possible and easy.  The great strength of DASCH is the ease at getting many hundreds of magnitudes, all from your home or office, with many of the logistic and technical problems solved.  A further strength of DASCH is that many stars can be examined for any mode of photometric behaviour, and this enables large scale search programs and large scale demographics programs.

Just as with CCD photometry (and all types of astronomical measures), DASCH has some issues that must be handled correctly, where abuse by inexperienced workers has lead to incorrect results.  The easiest trap (with the easiest solution) is to use the DASCH photometry calibrated with the `g' magnitudes of the {\it Kepler Input Catalog}, with this option being a historical relic from the earliest days of DASCH before APASS became their perfect calibration.  (Calibration with a photometric band far from the native Johnson B system leads to large errors in magnitudes, plus huge color terms that vary over time.)  The solution is simply to select the DASCH option to use the APASS calibration.  The next trap for the unwary is to use the DASCH magnitudes for plates where the image of some nearby crowding star touches the image of the target, in which case the DASCH photometry suffers large systematic errors.  Such crowding might occur preferentially for some telescopes (usually the patrol plates) or for some observing conditions (e.g., where focus or trailing works to make the star images touch).  The DASCH user must rigorously check for such conditions (best by looking at many of the scanned thumbnails available on the DASCH page, or possibly by rigorous use of the DASCH `A' quality flags), and ignore any questionable plates.  Further, I usually ignore any plate where the target star is within 0.3 mag, or so, of the plate limit, but this really depends on the needs of the science task at hand.  Even so, the DASCH light curves always have a small percentage of outliers.  (CCD surveys also have outliers, and my experience shows that the modern CCD light curves have a higher percentage of outliers.)  These outliers are caused by a wide array of plate defects, like from scratches on the emulsion, passing asteroids, double exposures, ghost images, and such.  But it is poor to simply toss out the outliers, as some might be from real variations on the star.  (For example, tossing out the eclipse plates of DQ Her and BT Mon as outliers would get rid of the primary signal being sought.)  So it is best to examine all outliers, one-by-one, with problematic plate defects often being obvious.  The moral is that DASCH users must examine a large number of plates by eye (or maybe by the A flags) for required quality control.

For the target CNe in this paper, I have exclusively used direct visual measures from the plates as compared to nearby APASS stars.  The majority of the plates have had independent examinations from 3--5 times.  This repetition is partly to beat down the measurement uncertainty, partly to provide a quantitative measure of the real error bars, and partly to form yet another experiment comparing the accuracy of the direct by-eye photometry versus the DASCH photometry.  Yet again, I find that the by-eye error bars are essentially equal to the 1-sigma error bars quoted by DASCH.  For BT Mon and DQ Her, the by-eye magnitudes are substantially better than for DASCH, because the target is always very close to the plate limit.

\section{DQ Her}

DQ Her is one of the most famous, best-observed, historical, and prototypical CNe.  The eruption started in 1934.953, reached a peak magnitude 1.6, took 100 days to fade by three magnitudes, underwent a deep dust dip (making it a D-class light curve), and it took over 32 years to decline to a quiescent level (Strope, Schaefer and Henden 2010).  DQ Her was the first nova to be discovered as an eclipsing binary (Walker 1956), and this realization opened the way to knowing the nature of nova systems and the cause of the nova eruptions.  In quiescence, DQ Her is the type-star for the class of Intermediate Polars, where the WD has a middle-strength magnetic field that forces the inner portions of the accretion disk into an accretion stream.  The quiescent magnitude is around 14.8, the eclipse has a variable depth from 1.05 to 2.43 mag in the B-band, the orbital period is 4.65 hours, and there is a 71-second oscillation (Patterson et al. 1978).

For DQ Her, the proposed $P_{pre}$=0.1932084 day value of Ahnert (1960) is wrong because he had too few plates that showed no eclipses, plus the computational abilities of the time were inadequate.  Further, about half his plates had exposure times comparable to half the orbital period, so no eclipses were possibly visible and these plates just added noise.  SP83 used Ahnert's magnitudes plus magnitudes from the Harvard plates, but they were not able to recover any $P_{pre}$ value in either DFTs or periodograms.  A few years ago, I realized that my set of Harvard plates was not complete (i.e., there were many more useable Harvard plates showing DQ Her before eruption) and the comparison sequence adopted from Gaposchkin (1956) was poor by modern standards.  So over the last decade, I have found all 34 useable Harvard plates from before the eruption, and measured the magnitude of DQ Her by comparison with the Johnson B magnitudes of nearby stars as provided by {\it APASS}.  Each of the Harvard plates has now been independently measured 3--5 times, for improving the measurement error.  Fortunately, with the 13 previously-unexamined plates, three plates showed DQ Her greatly fainter than its maximum light (i.e., in eclipse).  As shown below, the Ahnert and SP83 plates did not happen to show any eclipses, while the new finding of three eclipses makes for a confident measure of $P_{pre}$.  

\subsection{DQ Her Eclipse Times}

\begin{table}
	\centering
	\caption{DQ Her eclipse times}
	\begin{tabular}{llll} 
		\hline
		$T_i$	&   Eclipse minimum (HJD)   & Year & Source  \\
		\hline

$T_{-3}$	&	2420254.8215	$\pm$	0.003	&	1914.3322	&	MC 5493	\\
$T_{-2}$	&	2420606.8247	$\pm$	0.003	&	1915.2959	&	MC 8437	\\
$T_{-1}$	&	2425354.8176	$\pm$	0.008	&	1928.2952	&	RH 220	\\
$T_0$	&	2427786.1301	$\pm$	0.0002	&	1934.9518	&	$E_0$, Section 5.5	\\
$T_{1}$	&	2434954.7515	$\pm$	0.0003	&	1954.5797	&	Africano \& Olson	\\
$T_{2}$	&	2434954.9450	$\pm$	0.0003	&	1954.5803	&	Africano \& Olson	\\
$T_{3}$	&	2434955.7191	$\pm$	0.0003	&	1954.5824	&	Africano \& Olson	\\
$T_{4}$	&	2434983.7950	$\pm$	0.0003	&	1954.6593	&	Africano \& Olson	\\
$T_{5}$	&	2434984.7630	$\pm$	0.0003	&	1954.6619	&	Africano \& Olson	\\
$T_{6}$	&	2434985.7298	$\pm$	0.0003	&	1954.6646	&	Africano \& Olson	\\
$\ldots$	&	$\ldots$	&	$\ldots$	&	$\ldots$	\\
$T_{105}$	&	2450330.5724	$\pm$	0.0004	&	1996.6764	&	AAVSO (ZRE)	\\
$T_{106}$	&	2450632.4306	$\pm$	0.0003	&	1997.5029	&	Ogloza et al.	\\
$T_{107}$	&	2452413.5482	$\pm$	0.0003	&	2002.3793	&	Bianchini et al.	\\
$T_{108}$	&	2453943.7324	$\pm$	0.0001	&	2006.5687	&	AAVSO (COO)	\\
$T_{109}$	&	2453943.7328	$\pm$	0.0001	&	2006.5687	&	AAVSO (COO)	\\
$T_{110}$	&	2453948.7667	$\pm$	0.0001	&	2006.5825	&	AAVSO (COO)	\\
$T_{111}$	&	2453950.7031	$\pm$	0.0001	&	2006.5878	&	AAVSO (COO)	\\
$T_{112}$	&	2453950.8964	$\pm$	0.0001	&	2006.5884	&	AAVSO (COO)	\\
$T_{113}$	&	2453952.8334	$\pm$	0.0001	&	2006.5937	&	AAVSO (COO)	\\
$T_{114}$	&	2454313.5490	$\pm$	0.0001	&	2007.5812	&	AAVSO (VMT)	\\
$T_{115}$	&	2454318.3893	$\pm$	0.0001	&	2007.5945	&	AAVSO (VMT)	\\
$T_{116}$	&	2454379.3819	$\pm$	0.0001	&	2007.7615	&	AAVSO (VMT)	\\
$T_{117}$	&	2455012.9052	$\pm$	0.0001	&	2009.4960	&	AAVSO (GFB)	\\
$T_{118}$	&	2455014.8417	$\pm$	0.0001	&	2009.5013	&	AAVSO (GFB)	\\
$T_{119}$	&	2455065.7645	$\pm$	0.0001	&	2009.6407	&	AAVSO (KRV)	\\
$T_{120}$	&	2455066.7340	$\pm$	0.0001	&	2009.6434	&	AAVSO (KRV)	\\
$T_{121}$	&	2455440.4214	$\pm$	0.0001	&	2010.6665	&	AAVSO (MEV)	\\
$T_{122}$	&	2455441.3893	$\pm$	0.0001	&	2010.6691	&	AAVSO (MEV)	\\
$T_{123}$	&	2455831.7306	$\pm$	0.0001	&	2011.7378	&	AAVSO (SWIL)	\\
$T_{124}$	&	2455987.0153	$\pm$	0.0001	&	2012.1629	&	AAVSO (HMB)	\\
$T_{125}$	&	2456183.7330	$\pm$	0.0005	&	2012.7015	&	AAVSO (CMM)	\\
$T_{126}$	&	2456186.6368	$\pm$	0.0005	&	2012.7095	&	AAVSO (SWIL)	\\
$T_{127}$	&	2456191.6713	$\pm$	0.0001	&	2012.7233	&	AAVSO (SWIL)	\\
$T_{128}$	&	2456356.6372	$\pm$	0.0001	&	2013.1749	&	AAVSO (RJWB)	\\
$T_{129}$	&	2456847.4635	$\pm$	0.0001	&	2014.5187	&	AAVSO (CDZ)	\\
$T_{130}$	&	2456869.5369	$\pm$	0.0001	&	2014.5792	&	AAVSO (CDZ)	\\
$T_{131}$	&	2456870.5049	$\pm$	0.0001	&	2014.5818	&	AAVSO (CDZ)	\\
$T_{132}$	&	2457179.5240	$\pm$	0.0001	&	2015.4279	&	AAVSO (BPO)	\\
$T_{133}$	&	2457190.9473	$\pm$	0.0001	&	2015.4591	&	AAVSO (JJI)	\\
$T_{134}$	&	2457541.7891	$\pm$	0.0001	&	2016.4197	&	AAVSO (MZK)	\\
$T_{135}$	&	2457543.9186	$\pm$	0.0001	&	2016.4255	&	AAVSO (JJI)	\\
$T_{136}$	&	2457547.4040	$\pm$	0.0001	&	2016.4351	&	AAVSO (OYE)	\\
$T_{137}$	&	2457579.7385	$\pm$	0.0001	&	2016.5236	&	AAVSO (COO)	\\
$T_{138}$	&	2457906.3780	$\pm$	0.0001	&	2017.4179	&	AAVSO (MNIC)	\\
$T_{139}$	&	2457924.7716	$\pm$	0.0001	&	2017.4682	&	AAVSO (COO)	\\
$T_{140}$	&	2458202.8128	$\pm$	0.0001	&	2018.2295	&	AAVSO (DKS)	\\
$T_{141}$	&	2458225.8537	$\pm$	0.0001	&	2018.2925	&	AAVSO (DKS)	\\
$T_{142}$	&	2458234.7600	$\pm$	0.0001	&	2018.3169	&	AAVSO (DKS)	\\
$T_{143}$	&	2458239.7941	$\pm$	0.0001	&	2018.3307	&	AAVSO (DKS)	\\
$T_{144}$	&	2458335.6357	$\pm$	0.0001	&	2018.5931	&	AAVSO (DKS)	\\

		\hline
	\end{tabular}
\end{table}

DQ Her is an eclipsing binary and the literature contains many accurate measures of the time of deepest eclipse.  After the original discovery by Walker, many observers have reported times of mid-eclipse for DQ Her, with two big compilations in Africano \& Olson (1981) and Zhang et al. (1995).  The two compilations claim a measurement error of $\pm$30 seconds and $\pm$10 seconds respectively.  Further eclipse times in the literature are from Schoembs \& Rebhan (1989), Ogloza, Drozdz \& Zola (2000), and Bianchini et al. (2004).  These eclipse times are given in Table 1.  These are all heliocentric corrected times.  Each eclipse time is labeled with a notation $T_i$, where the `{\it i}' subscripts are consecutive integers within the table.  The table does not display the times $T_{7}$ to $T_{104}$ from 1956--1994, as these have already appeared collected in the literature (Africano \& Olson 1981; Zhang et al. 1995).

I have found published eclipse times only up until the year 2002, and it would be nice to extend this to the current year, as this can allow for a longer time interval to seek changes in $P_{post}$.  To fill this need, I have pulled well-sampled eclipse light curves from the {\it AAVSO} database.  For each, I have performed a chi-square fit for a parabola to the light curve near minimum.  This gives me 38 new eclipse times from 1996 and 2006--2018, along with the 1-sigma uncertainty.  In Table 1, the source for each time is designated with the {\it AAVSO} observer designation code in parentheses.

All of the above eclipse times have their reported 1-sigma measurement error.  But the total error in the minima times is substantially larger.  The systematic problem is that the system is flickering and varying randomly on all time scales, so variations between the ingress and egress branch will lead to a skewed minimum and a fitted parabola jerked to one side.  This is ubiquitous and inevitable, with nothing we can do to minimize it or to correct it.  So we have some systematic error that must be added in quadrature with the measurement errors so as to produce the total random error that must be used in any chi-square fits to the O-C diagram.

Table 1 also reports an eclipse time for the time of eruption, close to the year 1934.953.  This is designated $T_0$, and is a calculated quantity from the best joint fit (see Section 5.5).

Further, Table 1 also gives the three pre-nova eclipse times from the Harvard plates.  All three plates are limits on the magnitudes (i.e., DQ Her was not detected on the plate down to the stated magnitude), and the three exposure times for $T_{-3}$, $T_{-2}$, and $T_{-1}$ are 10, 10, and 68 minutes.  The listed time is for the center of the exposure. 

What is the formal uncertainty in the eclipse times from the archival plates?  The value to quote depends on whether we want the utter maximum possible deviation between the time of mid-exposure and the time of mid-eclipse, or whether we want the 1-sigma uncertainty on the eclipse time.  For the eclipse time analysis in Section 5.4.1, we want an ultraconservative maximum possible deviation.  For the two 10 minute exposures, the magnitude limit is so faint that the upper limit on the deviation is set mostly by the shape of the eclipse, where the recorded brightness can be dimmed by $>$0.8 mag or more only if the plate mid-exposure time is within 7--8 minutes of mid-eclipse, even with the recent very-deep eclipses.  So $T_{-3}$ and $T_{-2}$ have maximum conceivable errors of $\pm$0.005 d for purposes of Section 5.4.1.  For the $T_{-1}$ plate with the 68 minute exposure, almost all of the duration of the eclipse must be during the plate exposure, as this is required to produce the observed magnitude drop from maximum light.  (See equations 3 and 4 of SP83.)  For a 33 minute total eclipse duration, the time of mid-eclipse must be more than roughly 15 minutes from the start or end time of the plate exposure, which is to say that the difference between the mid-eclipse and mid-exposure time must be 19 minutes or less.  So for the uncertainty as needed in Section 5.4.1, $T_{-1}$ has a maximal deviation of 0.013 d.

The formal error on the eclipse times is the 1-sigma interval , which is taken as the standard deviation over the probability distribution.  For the 68 minute exposure, the probability is uniform for the mid eclipse occurring within an interval from -19 to +19 minutes after mid-exposure.  The 1-sigma value is 0.6 times 19 minutes, or 11 minutes.  For the 10 minute exposures, the probability is uniform from -7.5 to +7.5 minutes, for a 1-sigma error bar of 4.5 minutes.  These error bars are too small to be usefully displayed in Figure 3.  These error bars are tabulated in Table 1 and plotted in Figure 4.

\subsection{DQ Her Post-Eruption Orbital Period}

\begin{figure}
	\includegraphics[width=\columnwidth]{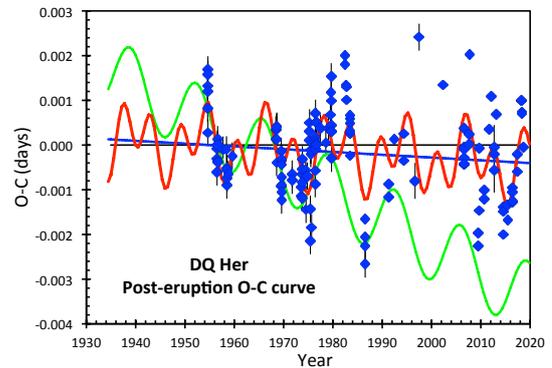}
    \caption{$O-C$ curve for post-eruption eclipse times.  The $O-C$ values are the differences between the observed eclipse times (see Table 1) and the eclipse times predicted by a linear model of Ogloza et al. (2000) with $T_i$=2434954.9438+N$\times$0.193620901.  We see that the measures within the few months of each observing season are more scattered than given by their formal measurement, and I attribute this to the usual flickering of DQ Her on the ingress and egress making for a ubiquitous jitter.  The seasonal averages vary around much more than their uncertainty, demonstrating that there is some unknown year-to-year variation at the $\lesssim$0.001 day level.  The old model of Ogloza et al. (2000) is shown with the flat black line.  The best fit line of all 144 post-eruption eclipse times is shown with the deep-blue line with a small slope.  The green single sinewave is that from Patterson et al. (1978), which was only intended as a description for the 1954--1977 eclipse times, and this model with periodicity soon stopped being an effective model of the $O-C$ curve.  In a further attempt to describe the 144 eclipse times, the red wiggly line is the best fit model with two sinewaves, each period chosen for the two highest peaks in the DFT.  All of these models fails to follow the observations closely, hence demonstrating that the extrapolations back to the date of the eruption (1934.95) must have a substantial uncertainty at the 0.001 day level.}
\end{figure}

I have 144 eclipse times from 1954--2018 in Table 1, and these can be used to establish $P_{post}$.  With the linear ephemeris of Ogloza et al. (2000), $T=2434954.9438+N\times0.193620901$, I have constructed an $O-C$ diagram in Fig. 1.  This model is represented as a flat line in the figure.  Extrapolated back to the start of the nova, there is an eclipse time at HJD 2427786.1299.

The $O-C$ diagram shows substantial scatter, far above the nominal measurement errors given in Table 1.  What we see is that the values within the few months of each observing season are tightly clustered (yet still more scattered than their error bars), while the yearly averages move around greatly more than the uncertainty in the yearly averages.  I take the extra scatter inside each observing season to be the result of the ordinary flickering causing an extra random jitter.  For the 18 years with four or more eclipse times, the median RMS in the $O-C$ is 0.0004 days, which is always much larger than the measurement error.  So fits to the $O-C$ diagram must have this additional random systematic error added in quadrature to the quoted measurement errors.

The year-to-year variations are highly significant.  But they appear to be aperiodic (with a time scale of a few years) and to not have any apparent long-term trend.  The cause of these variations is not known.  Various attempts have been made to describe these variations with single or double sinewaves, all with little success.  Critically, for the task at hand (getting $P_{post}$ at the time just after the eruption in 1939.94) these variations in the O-C curve are all small.  The RMS scatter about the zero line in the $O-C$ curve is 0.0009 days (1.3 minutes), so the epoch at the time of the eruption ($E_0$) can have an uncertainty of this size.

I have fitted a linear ephemeris to this 1954--2018 set of eclipse times.  I get a period of 0.1936208981$\pm$0.0000000019 days and an epoch of HJD 2445500.89326$\pm$0.00008.  Extrapolated back to the time of the eruption, there is an eclipse time at HJD 2427786.1301.  This is represented by the tilted line in Fig. 1.  There is no significant $\dot{P}$ term for a steady period change, with the rather severe limit of $|\dot{P}|$$<$5$\times$10$^{-13}$ days/cycle.

I have plotted a single sinewave on top of the linear ephemeris, with a period of 4900 days, as used by Patterson et al. (1978) to describe their early set of eclipse times (red sinewave in Fig. 1).  With their particular ephemeris, the model diverges increasingly after the date of their last observation, being now 0.0026 days too low.  The 4900 day time scale has not persisted past the 1990's.  So it is clear that the apparent 4900 day sinewave is just apparent because the then-available eclipse times happened to have two minima in the $O-C$ curve around 1959 or later plus a vague minimum around 1973.  The non-existence of the 4900 day cycle is no surprise to anyone.

A Fourier transform of the post-eruption $O-C$ curve shows no significant or prominent peaks.  Still, as a description of the variability, I have constructed a model with the periods of the two highest peaks, at 4950 days (describing the pre-1978 variability, just as in Patterson et al. 1978) and 2116 days.  The best fit amplitudes are 0.00046 days and 0.00059 days respectively.  The resultant model is displayed in Fig. 1 as the complex double sinewave.  Just to be sure that it is stated, this double sinewave is merely an attempt at a description of the $O-C$ curve, and there is neither any expectation that the periods are physical, nor that the behavior outside the interval 1954--2018 will follow the extrapolated model.

These four models span the typical results for extrapolating the observed $O-C$ curve back to 1939.953.  As expected, we see that the epoch $E_0$ is uncertain by approximately 0.0010 days (1.4 minutes), due to the year-to-year variations, with the linear ephemerides providing the central value of 2427786.1301.  The $P_{post}$ values for the linear ephemeris is close to 0.1936209898 days as from the best linear fit with an uncertainty of something like $\pm$ 0.00000026 days or so.  That is, with the intrinsic variations in period during quiescence, the period at the date of the eruption has an uncertainty of order 1.3 ppm.

\subsection{DQ Her Pre-Eruption Magnitudes}

\begin{table}
	\centering
	\caption{DQ Her pre-eruption B magnitudes}
	\begin{tabular}{llll} 
		\hline
		Mid-exposure (HJD)	&   Year   & B (mag) & Source  \\
		\hline

2412902.9094	&	1894.2037	&	15.22	&	A365	\\
2412963.8372	&	1894.3705	&	14.68	&	I11106	\\
2414569.5719	&	1898.7668	&	15.22	&	I21340	\\
2415150.7092	&	1900.3579	&	14.89	&	I25320	\\
2415186.7886	&	1900.4566	&	15.03	&	I25459	\\
2415281.5987	&	1900.7162	&	14.70	&	I25810	\\
2415593.6225	&	1901.5705	&	14.82	&	A5497	\\
2415941.5896	&	1902.5232	&	15.05	&	I28978	\\
2415985.6194	&	1902.6437	&	14.72	&	I29143	\\
2419583.5041	&	1912.4942	&	15.02	&	MC1881	\\
2420254.8215	&	1914.3322	&	>15.8	&	MC5493	\\
2420393.5326	&	1914.7119	&	15.23	&	MC6335	\\
2420606.8247	&	1915.2959	&	>15.9	&	MC8437	\\
2421019.7939	&	1916.4265	&	14.69	&	MC10833	\\
2421105.5598	&	1916.6614	&	14.93	&	MC11073	\\
2423176.8370	&	1922.3322	&	15.05	&	I41208	\\
2423176.8464	&	1922.3322	&	14.80	&	I41209	\\
2423190.8151	&	1922.3705	&	14.93	&	I41255	\\
2423295.5019	&	1922.6571	&	15.02	&	MC18945	\\
2423528.8127	&	1923.2959	&	15.11	&	MC19723	\\
2425350.8230	&	1928.2843	&	15.05	&	RH206	\\
2425354.8176	&	1928.2952	&	$\gtrsim$15.6	&	RH220	\\
2425408.7241	&	1928.4428	&	14.90	&	RH313	\\
2425800.6650	&	1929.5159	&	15.06	&	MC24335	\\
2426744.6515	&	1932.1003	&	14.72	&	Ahnert (948)	\\
2426750.6557	&	1932.1168	&	14.75	&	Ahnert (959)	\\
2426767.5702	&	1932.1631	&	14.84	&	Ahnert (969)	\\
2426771.5453	&	1932.1740	&	14.93	&	Ahnert (981)	\\
2426827.4772	&	1932.3271	&	15.28	&	Ahnert (1008)	\\
2427131.6121	&	1933.1598	&	14.90	&	Ahnert (1138)	\\
2427133.6592	&	1933.1654	&	15.11	&	Ahnert (1147)	\\
2427153.4999	&	1933.2197	&	15.16	&	Ahnert (1151)	\\
2427154.5149	&	1933.2225	&	15.23	&	Ahnert (1154)	\\
2427155.5880	&	1933.2254	&	14.99	&	Ahnert (1158)	\\
2427158.5561	&	1933.2336	&	14.87	&	Ahnert (1169)	\\
2427181.4669	&	1933.2963	&	14.93	&	Ahnert (1181)	\\
2427192.8157	&	1933.3274	&	15.14	&	RH5078	\\
2427211.7685	&	1933.3792	&	14.98	&	RH5119	\\
2427462.6712	&	1934.0662	&	15.28	&	Ahnert (1386)	\\
2427482.6397	&	1934.1208	&	15.22	&	Ahnert (1402)	\\
2427483.6537	&	1934.1236	&	14.85	&	Ahnert (1407)	\\
2427511.5416	&	1934.2000	&	14.91	&	Ahnert (1418)	\\
2427513.5587	&	1934.2055	&	15.20	&	Ahnert (1421)	\\
2427519.8729	&	1934.2228	&	14.96	&	RH5839	\\
2427535.8379	&	1934.2665	&	14.94	&	RH5898	\\
2427657.6761	&	1934.6001	&	15.16	&	MC27364	\\
2427661.6066	&	1934.6108	&	15.05	&	RH6135	\\
2427686.5933	&	1934.6792	&	15.00	&	IR100	\\
2427694.5567	&	1934.7010	&	15.19	&	RH6190	\\
2427712.5219	&	1934.7502	&	14.90	&	RH6208	\\
2427715.5281	&	1934.7585	&	14.89	&	RH6222	\\
2427745.2578	&	1934.8399	&	14.97	&	Ahnert (1615)	\\

		\hline
	\end{tabular}
\end{table}

The Harvard and Sonneberg collections of archival sky photographs provide the only sources of magnitudes before 1934.953.  The Harvard magnitudes have been previously reported  by Gaposchkin (1956) and SP83, while the Sonneberg magnitudes comes from Ahnert (1960).  Both of these light curves are on a non-standard system, due to the adopted comparison sequence having the usual differences, for the times, from the B-system.  The adopted magnitudes for each comparison star have been given, so it is easy to convert these non-standard magnitudes into the Johnson B system with the procedure in Johnson et al. (2014).  For the Sonneberg plates, all I have are Ahnert's converted magnitudes (see Table 2), and this should be adequate.

For the Harvard plates, I have made three independent measures of most of the magnitudes by direct visual examination of the plates.  These measures were made widely separated in time, with the plates being in essentially a random order and with an unknown prior-magnitude before each measure.  By many widely varied experiments over the years, the visual measures by an experienced observer of ordinary plates with good comparison sequences yields real 1-sigma error that are equal to that produced by automated programs (such as DASCH).  For example, for the 16 DQ Her plates with DASCH magnitudes, the comparison with my visual measures shows an average difference of 0.009 mag and an RMS scatter in the differences equal to the average of the quoted DASCH error bars.  A fourth set of measures was to use the DASCH photometry for the same plates.  A fifth set of magnitudes was to use a visual examination of the scanned thumbnails of the plates, as available from DASCH.  All of the sets are consistent with zero offsets, and the individual magnitudes only show the expected scatter from measurement errors.  Each of the plates has three to five measures of the DQ Her brightness.  I have straight averaged these measures to obtain the final magnitude (see Table 2).  

Substantial problems arise because one Harvard plate (RH6011) and all the early Sonneberg plates have exposure time of 1.9 to 3.0 hours.  For comparison, the orbital period is 4.65 hours, while the FWHM of the eclipse is near a quarter-hour.  One trouble is that for a long exposure plate, the drop in the total illumination will be small.  For an exposure of 120 minutes, roughly 8$\times$ the eclipse FWHM, the received light will only be $7/8$ of that without any eclipse, for a drop in illumination corresponding to 0.14 mag.  For the one-sigma measurement error of 0.17 mag (see below), no two-hour plate can show a significant eclipse.  There are many more fundamental and practical problems with the use of these very long exposure plates, as they provide virtually zero information at the cost of large added noise.  The best procedure is to simply not use these magnitudes, so I have not even included them into Table 2 or subsequent analysis. 

In the end, I have 52 pre-eruption B magnitudes for DQ Her.  This includes 18 magnitudes converted to B band from Ahnert (1960) plus 34 B magnitudes from the Harvard plates.  This includes three plates where DQ Her appears substantially fainter than all other plates, and these are plates that cover old eclipses.

\subsection{DQ Her Pre-Eruption Orbital Period}

I have determined $P_{pre}$, using the pre-eruption light curve alone, with three different methods:

\subsubsection{Eclipse Times}

There are only three pre-eruption eclipse times in Table 1, and this might seem inadequate.  Actually, we can get away with two or perhaps even one eclipse plate and still derive a highly confident $P_{pre}$.  (This was exhaustively proven for the classical nova QZ Aur, in a virtually identical case, see Schaefer et al. 2019.)  We have two more valuable sources of information available.  First, the at-maximum magnitudes provide a strong constraint that there must be none of these with a phase inside the eclipse duration.  This is critical for eliminating the various possible false alarm periods.  Second, the post-eruption $E_0$ value is like an eclipse time at the very end of the pre-eruption period, and it can serve as a fourth eclipse.  In all, we have good redundancy and will produce a very highly confident $P_{pre}$.

My first method for is to look at the time differences between the four eclipse times ($T_{-3}$, $T_{-2}$, $T_{-1}$, and $T_0=E_0$).  The eclipse period will be $(T_i-T_j)/\Delta N_{ij}$, where $\Delta N_{ij}$ is the cycle count between the eclipses.  This uses the point that $\dot{P}$ is very small, as seen in the post-eruption data.  The total time interval is just over 20 years, so the period will be measured with very high accuracy.  The trick is to determine the cycle count.  Here, I will only look for $P_{pre}$ values that are within 3000 ppm of $P_{post}$, as it is hard physically to get a larger period change.  Within this constraint, the task is to make an exhaustive search for a period that reproduces all the eclipse time intervals.

To start, the shortest time interval is $T_{-2}-T_{-3}$=352.0032$\pm$0.0042 days.  Only three values of $\Delta N_{-2,-3}$ have periods close enough to $P_{post}$, with values of 1817, 1818, and 1819.  These produce periods of 0.1937277$\pm$0.0000023, 0.1936211$\pm$0.0000023, and 0.1935147$\pm$0.0000023 days.  Our attention is immediately focused on the middle period, as it very near to  $P_{post}$.  Further, the 1817 and 1819 possibilities are certainly rejected because several at-maximum magnitudes are present for phase values close to those of the pair of eclipse plates.

The next longest time interval is $T_0-T_{-1}$=2431.3126$\pm$0.0120 days.  Within the physically possible search range, we have possible values of $\Delta N_{0,-1}$ from 12545 to 12569.  But only two of these cases are consistent with the possibilities from the previous paragraph.  This is for $\Delta N_{0,-1}$=12557 with $P_{pre}$=0.19362209$\pm$0.00000096, while the second is for $\Delta N_{0,-1}$=12564 with $P_{pre}$=0.19351421$\pm$0.00000096.  The first is a close match to the $\Delta N_{-2,-3}$=1818 period, which is the one so close to $P_{post}$.  Again, the second one is rejected because it produces a folded light curve with a few points with $B\approx$15.0 inside the eclipse interval.  So we already have our unique answer, with just two eclipse time intervals, even without resorting to the at-maximum light curve.

The next longest interval between eclipses can be used to improve the accuracy of the measured $P_{pre}$.  For $T_{-1}-T_{-2}$=4747.9928$\pm$0.0124 days, the cycle count must be between 24498 and 24546.  But of these possibilities, only $\Delta N_{-1,-2}$=24522 produces a period that is consistent with any of the possibilities from $\Delta N_{-2,-3}$ and $\Delta N_{0,-1}$.  So again, we can be sure that we have made the correct cycle counts, and we have a confident period.  With this, $P_{pre}$=0.19362176$\pm$0.00000050.  

The most accurate $P_{pre}$ comes from the longest time interval between eclipses of $T_0-T_{-3}$=7531.3086$\pm$0.0031 days.  The calculated period remains close enough to $P_{post}$ for $\Delta N_{0,-3}$ values from 38859 to 38936.  Again, we get a match between all the candidate periods only for one case, with $\Delta N_{0,-3}$=38897.  With this, we have $P_{pre}$=0.19362184$\pm$0.0000008.

I am going through this analysis because anyone can see that the derived period is highly confident, unique, and very accurate.  All this with a simple analysis involving only subtraction and division.

\subsubsection{Periodogram}

\begin{figure}
	\includegraphics[width=\columnwidth]{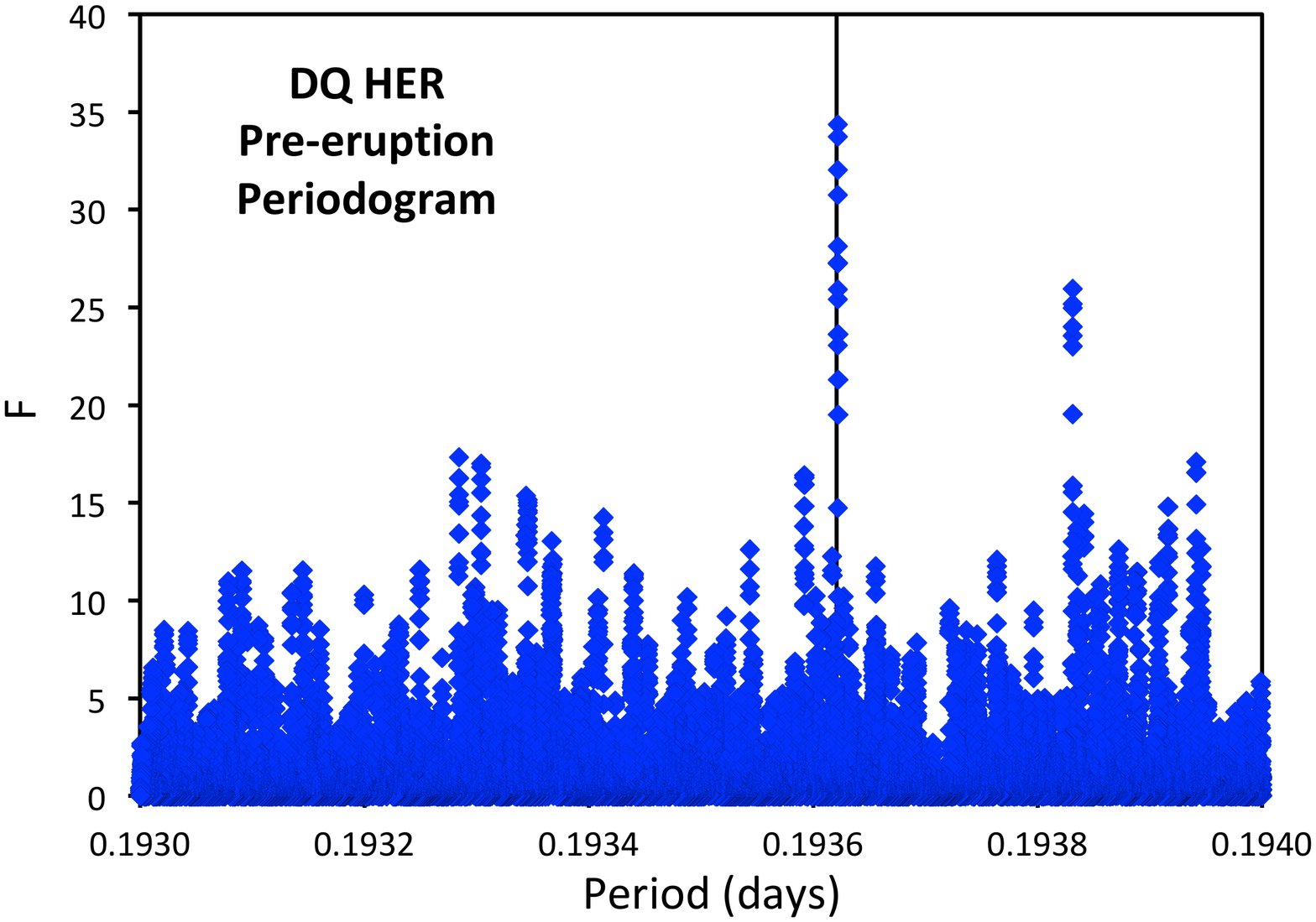}
	\includegraphics[width=\columnwidth]{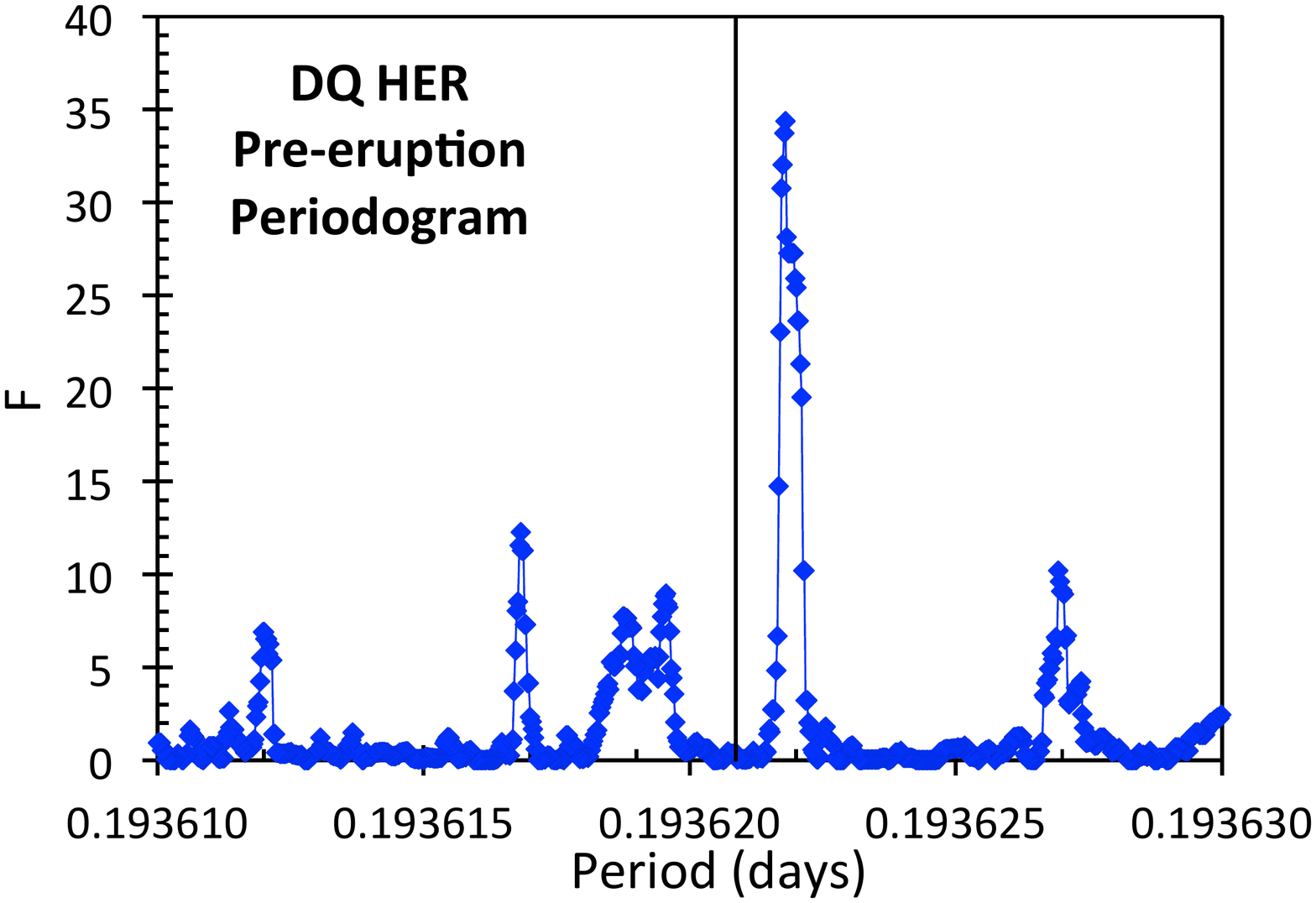}
    \caption{Periodogram for pre-eruption magnitudes of DQ Her.  The blatant peak is at a period of 0.193621801 days, with this being highly significant.  The post-eruption period is indicated by the vertical black line.  The top plot shows the entire range of trial periods (0.193--0.194 d).  Far below the highest peak, we see an alias period reaching up to 25 and several noise peaks reaching up to 17. The bottom plot shows a blow up around $P_{post}$ to show that the orbital period has {\it decreased} across the 1934 nova eruption.}
\end{figure}

A better analysis will be to calculate a periodogram.  The type of periodogram that I have used is one where the light curve is folded on many trial periods, and a figure-of-merit is calculated for each trial period, and a plot of these will show the true period as an isolated peak well above the noise.  I have calculated the figure-of-merit for trial periods from 0.193 to 0.194 days (from -3200 to +2000 ppm difference from $P_{post}$) with 39,640 trial periods evenly spaced in period yielding a greatly over-sampled plot.  (There are roughly 2000 independent trial periods examined in this whole range.)  The figure-of-merit chosen is the F-test value for comparing the model with a light curve that is flat versus a model with a realistic model for the shape of the light curve of the eclipsing binary (as taken from modern photometric time series).  When the trial period is very close to the true period, the folded light curve will differ greatly from a flat line, resulting in a large statistic.  When the trial period is far from the true period, the eclipse plates will spread around in phase resulting in a light curve that looks flat, resulting in a near zero value for the statistic.  This periodogram uses the knowledge from $E_0$ so as to know the phase at which an eclipse must occur for the trial period.  (Again, the $E_0$ value is known from the post-eruption $O-C$ curve in Fig. 1 has an uncertainty of $\pm$0.001 d, and this is more than good enough to make for an unambiguous periodogram.)  The periodogram also uses in full the information from the at-maximum magnitudes, as a trial period with at-maximum magnitudes occurring within the phase range for the duration of the eclipse will provide a severe penalty in the figure-of-merit.  This method has the strong positive trait that it exhaustively checks all possible trial periods over a very wide range.

The full periodogram shows only one peak (at 0.193621801 d) at the highly significant level of 34.4.  The second highest peak is an alias, where only two eclipse plates line up, and one at-maximum plate falls in the center of the eclipse duration.  The next highest peak rises to 17.3, with four more noise peaks higher than 15.0.  The highest peak is very significant, despite the large number of trial periods examined.  So we have our answer for $P_{pre}$.

Fig. 2 shows a small section of the full periodogram plot, just from trial periods 0.19361 to 0.19363 days.  We see the high and isolated peak at 0.193621801 days.  The vertical line shows the position of $P_{post}$, and the peak is blatantly different with a longer orbital period.  That is, we can see that the orbital period of DQ Her has {\it decreased} across its 1934 eruption.

\subsubsection{Chi-square Fit to an Eclipse Profile}

\begin{figure}
	\includegraphics[width=\columnwidth]{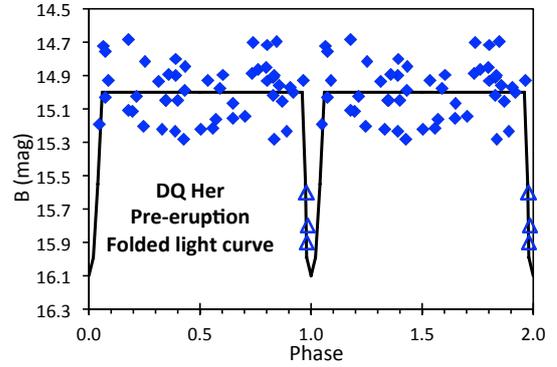}
    \caption{Folded light curve for pre-eruption magnitudes of DQ Her.  The 52 pre-eruption magnitudes are here folded on the best fit ephemeris with $P_{pre}$=0.193621775 days.  Each datum is plotted twice here, once for the phase and once for one-plus-the-phase.  The light curve model, showing the adopted eclipse profile for the chi-square fits, is the black curve that is flat from phases 0.06--0.96 and 1.06--1.96.  The at-maximum magnitudes are all in the flat section, with an average of B=15.00 and an RMS scatter of 0.17 mag.  This means that none of the at-maximum magnitudes are inside the duration of the eclipse.  The three triangles during the eclipse are limits.  The point of showing this folded light curve is to see that the pre-eruption ephemeris results in a perfect division, with no at-maximum points inside the eclipse phase, so as to create a gap, and all the in-eclipse plates being inside the eclipse duration.  This provides a quick visual confirmation that pre-eruption periodicity is significant and real, and that the $P_{pre}$ value is correct.}
\end{figure}

The third method to measure $P_{pre}$ is to make a chi-square fit of the magnitudes against a model of the eclipsing binary light curve shape.  This has the strong advantage over the periodogram of providing a ready means of measuring $E_0$, providing a well-known mechanism for calculating the exact error bars on $P_{pre}$, and providing a well-known means for calculating the significance of the periodicity.

The chi-square calculation requires a 1-sigma error bar for each magnitude.  With my extensive experience at quantitative analysis of the old plates, I would estimate an average error bar of around $\pm$0.20 mag, with this being relatively large due to DQ Her often being near the plate limiting magnitudes.  With essentially identical plates and conditions, for QZ Aur, I measured an average error bar of $\pm$0.18 mag.  For 16 of the specific DQ Her plates, DASCH provides individual error bars, with their average being $\pm$0.20 mag.  The best way to get the average error bar specific for DQ Her is to look at the RMS scatter of the 41 magnitudes which have a phase from 0.10 to 0.90, and that is $\pm$0.17 mag.   These out-of-eclipse plates have an average magnitude of B=15.00.

There is no perfect model for the eclipse profile.  One trouble is that the star flickers fast, making each profile vary substantially, with a different profile from eclipse-to-eclipse.  Another trouble is that the eclipse amplitude is changing substantially over the decades after the nova (Patterson et al.1978; Zhang et al. 1995), so we have no confident idea as to the eclipse amplitude before the nova.  The depth of the eclipse does not matter for the analysis, because the only deep magnitudes are {\it limits} near phase zero.  For the model, I have adopted an eclipse depth of 1.1 mag, although it could easily be 2.0 or 3.0 mag.  The total duration and the FWHM duration should both be similar to the modern values.  I have adopted a FWHM of 0.03 in phase, with the eclipse lasting from -0.04 to +0.06 in phase, so as to represent the egress as being slower than the ingress.  The model eclipse profile is displayed in Fig. 3.

The chi-square as a function of trial period shows a blatant minimum (greatly lower than any other noise feature) centered at $P_{pre}$=0.193621775$\pm$0.000000010 days.  (The formal 1-sigma error bar is somewhat asymmetric, so I am quoting the value towards $P_{post}$.)  This best fit has $E_0$=2427786.1304$\pm$0.0004.  The chi-square equals 46.8, for 49 degrees of freedom.

The pre-eruption $E_0$ is completely independent of the post-eruption $E_0$, yet it differs by only 0.0003$\pm$0.0011 days.  This close of a coincidence is improbable at the part-per-thousand level for the case that the pre-eruption period is due to some artifact or chance alignments.  With this, we have a strong argument that the pre-eruption orbital period is real and significant.

Another way to test the significance of the discovered pre-eruption periodicity is to perform an F-test comparing the above best fit model with the zero-amplitude model.  That is, how much does the chi-square decrease by just changing the one amplitude parameter from 0.0 to 1.1 mag.  (In the zero-amplitude model, the constant magnitude shifts to B=15.04 with the inclusion of the near-eclipse phases.)  The zero-amplitude model produces a chi-square of 122.8.  That is, by only changing one model parameter (the amplitude), the chi-square was decreased by 76.0.  The probability of this large improvement in chi-square by merely changing the amplitude is 1.6$\times$10$^{-6}$ for data without the periodicity being real.

A third way to test the significance of the $P_{pre}$ is with an F-test comparing the best fit model versus an identical model where the period is allowed to vary.  For $P_{pre}$ varying over the search range, the median chi-square is 220, which leads to a probability of 1.4$\times$10$^{-7}$ for data with no true periodicity to produce such a low chi-square value of 46.8.  With the 2000 trial periods examined, the $P_{pre}$ value is significant at the 0.00028 probability level.

In all analyses, the existence and value of $P_{pre}$ is highly significant.  So we can be very confident that the pre-eruption orbital period has $P_{pre}$=0.193621775$\pm$0.000000010 days and $E_0$=2427786.1304$\pm$0.0004.  The phase curve for the best fit is displayed in Fig. 3.  Critically, we see a time interval at zero phase with all three deep eclipse plates well centered, zero at-maximum plates within the interval of the eclipse duration, and all the at-maximum magnitudes forming a nice flat light curve with no exceptions.

\subsection{DQ Her Joint Fit}

So we have highly confident measures of $P_{pre}$ and $P_{post}$, each with consistent $E_0$ values.  But the physics of the situation tells us that the $O-C$ curve must be continuous everywhere, which is to say that the stars do not jump forwards or backwards in their orbits.  So formally, we need to have one joint fit where $E_0$ is held to be equal for both pre-eruption and post-eruption segments.  This will provide the overall best and final value for $\Delta$P.  For the case of DQ Her, where the two $E_0$ are so close, the joint fit will make for only small changes.

The joint fit is to minimize the sum of the chi-squares for the 52 pre-eruption magnitudes fitted to the model eclipse light curve folded on an ephemeris of $E_0 + N\times P_{pre}$ plus the chi-square of the 144 post-eruption eclipse times fitted to a model with eclipses at $E_0 + N\times P_{post}$.  This joint fit has 52+144=196 data points and a three parameter model for 193 degrees of freedom.

The joint fit has an isolated and very low minimum for one period, with a chi-square far smaller than any minimum at a other period.  I get $P_{pre}$=0.1936217610$\pm$0.0000000055 days, $P_{post}$=0.1936208977$\pm$0.0000000017 days and $E_0$=2427786.1301$\pm$0.0002.  This gives a change in orbital period across the nova eruption of $\Delta$P=-0.000000863$\pm$0.0000000058, or $\Delta$P/P=$-$4.46$\pm$0.03 ppm change.  If a $\dot{P}$ term is added to this joint fit, I get a tight limit of $|\dot{P}|$$<$2$\times$10$^{-13}$ days/cycle.

\subsection{$\Delta$P for DQ Her}

Fig. 4 presents an $O-C$ diagram for the 144 post-eruption eclipse times plus the 3 pre-eruption eclipse times (as tabulated in Table 1), all against the best fit post-eruption linear ephemeris.  We see that the year-to-year jitter in the eclipse times is negligibly small as compared to the kink at the time of the eruption.  We also see that the pre-eruption $O-C$ values are small compared to the period, and this is good evidence that we have the cycle counts correct.  We see that the pre-eruption eclipse plates are certainly not along an extension of the post-eruption line.  This demonstrates that there is a period change.  That the kink at 1934.95 is downward shows that the orbital period {\it decreased} across the eruption.

So, we have a highly confident and robust measure of the period change for the DQ Her eruption.  The period {\it decreased}, with $\Delta$P/P=$-$4.46$\pm$0.03 ppm.

The $\Delta$P for DQ Her is negative, and this cannot be by any of the usual mechanisms of mass ejection, FAML, or magnetic braking of the normal star in the nova ejecta.  The only known mechanism for getting a negative $\Delta$P is with $\Delta$P$_{jet}$.  But with DQ Her changing its period by only 4.46 ppm, the jet asymmetry must be very small, undetectably small.  Indeed, the DQ Her nova shell has no noticeable monopolar asymmetry.  (The shell is elliptical in shape, but such will not create any net jetting.)  Asymmetries at the part-per-million level are not detectable.

\begin{figure}
	\includegraphics[width=\columnwidth]{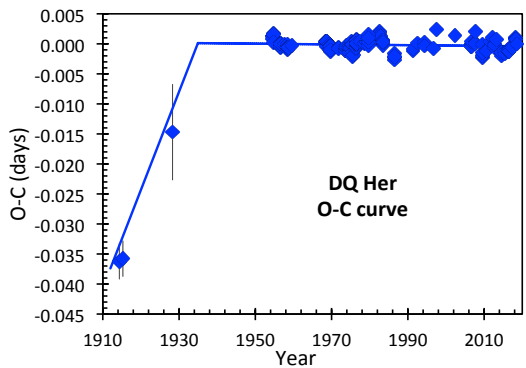}
    \caption{$O-C$ curve for DQ Her.  This $O-C$ curve is the same as for Fig. 1, except that the three pre-eruption eclipse times ($T_{-3}$, $T_{-2}$, and $T_{-1}$) are included.  (In principle, each point could be plotted one or more orbital periods higher or lower, but in practice, such cycle-count problems have been disproved by all three analyses in Section 5.4.  In this Figure, this is quickly seen visually as the pre-eruption and post-eruption $O-C$ curve must connect continuously with no vertical jumps, whereas if either branch, or any individual point, is raised or lower by multiples of 0.1936 d, then there must be discontinuities where the $O-C$ curve has very large vertical jumps.)  The broken line is from the joint best fit based on 52 pre-eruption magnitudes plus 144 post-eruption eclipse times, with the break being at the date of the nova eruption in 1934.953.  The best fit pre-eruption segment is slightly offset above the three measured eclipse times because the chi-square fit is also optimizing on placing a gap in the at-maximum magnitudes centered on the central eclipse times.  Importantly, the year-to-year variations for the post-eruption $O-C$ curve are seen to be small when compared to the period change across the eruption.  Also importantly, the pre-eruption $O-C$ values are all small compared to the orbital period, showing that it is easy to keep the correct cycle count between eclipses.  This figure shows that the change in the orbital period is highly significant.  The point of this figure is that it is certain that the orbital period ${\it decreased}$ across the nova event.}
\end{figure}

\section{BT Mon}

BT Mon (Strope, Schaefer \& Henden 2010) started its slow-moving eruption in 1939.689, was discovered on the Harvard plates, and peaked at 8.1 mag.  The eruption light curve shows a very flat top, with the star being constant in magnitude for at least 75 days.  As such, this is a prototype of the F-class novae, for which the cause of the flat-top is still unknown.  In quiescence, the star is around 15.7 mag, while Robinson, Nather, \& Kepler (1982) discovered deep eclipses with a period very close to one-third of a day.

SP83 used 68 pre-eruption plates to find seven plates in eclipse and measure $P_{pre}$.  The derived $\Delta$P /P is +39$\pm$4.8 ppm.  From 1983 until 2017, this was the only published value for any CN.  And soon after 1983, this measure that CNe having their orbital separation {\it increasing}, became part of the original motivation for the Hibernation Model.  

SP83 did a good job, but I can make now make improvements that were not possible back in 1983:  (1) I have collected many new post-eruption eclipse times from Steward Observatory, from time series reported to the AAVSO, and from the Harvard plates, all giving an accurate time history of $P_{post}$ and the steady $\dot{P}$ from 1941 to 2018, so as to get the very best $P_{post}$ for the time just after the eruption.  (2) I have used the modern comparison sequence from APASS so that the magnitudes are correctly placed into the Johnson B system.  (3) I have exhaustively searched the Harvard collection for useful plates, and I have collected more magnitudes from the Maria Mitchell plates, so I now have a total of 90 pre-eruption magnitudes.  (4) I have remeasured the Harvard plates 3--5 times independently, so as to beat down the measurement error.  (5) I have analyzed the pre-eruption light curve with the definitive chi-square analysis fitting to the modern eclipsing binary light curve, and performed a joint fit with the pre- and post-eruption data.  This is all a lot of work, but this was done as a test of the prior result, as a demonstration of the methods in this paper, and as an improvement in the accuracy of the measured $\Delta$P/P.

\subsection{BT Mon Eclipse Times}

\begin{table}
	\centering
	\caption{BT Mon eclipse times}
	\begin{tabular}{llll} 
		\hline
		$T_i$	&   Eclipse minimum (HJD)   & Year & Source  \\
		\hline
$T_{-10}$	&	2420435.9047	$\pm$	0.0300	&	1914.824	&	HCO (MC 6746)	\\
$T_{-9}$	&	2420451.8595	$\pm$	0.0300	&	1914.868	&	HCO (MC 6920)	\\
$T_{-8}$	&	2425943.5590	$\pm$	0.0100	&	1929.907	&	HCO (MC 13775)	\\
$T_{-7}$	&	2426035.3426	$\pm$	0.0200	&	1930.158	&	HCO (MF 13964)	\\
$T_{-6}$	&	2427397.9279	$\pm$	0.0100	&	1933.889	&	MMO (NA1603)	\\
$T_{-5}$	&	2427425.6971	$\pm$	0.0300	&	1933.965	&	MMO (NA1616)	\\
$T_{-4}$	&	2429340.3113	$\pm$	0.0200	&	1939.207	&	Wachmann (1968)	\\
$T_{-3}$	&	2429344.3210	$\pm$	0.0100	&	1939.218	&	Wachmann (1968)	\\
$T_{-2}$	&	2429364.3492	$\pm$	0.0200	&	1939.273	&	Wachmann (1968)	\\
$T_{-1}$	&	2429365.3491	$\pm$	0.0200	&	1939.275	&	Wachmann (1968)	\\
$T_{0}$	&	2443491.7159	$\pm$	0.0004	&	1939.689	&	$E_0$	\\
$T_{1}$	&	2430069.3732	$\pm$	0.0050	&	1941.203	&	1941 (5 eclipse plates)	\\
$T_{2}$	&	2432879.4340	$\pm$	0.0167	&	1948.896	&	1948 (1 eclipse plate)	\\
$T_{3}$	&	2433705.3048	$\pm$	0.0100	&	1951.158	&	1950--1951 (2 eclipse plates)	\\
$T_{4}$	&	2434440.3415	$\pm$	0.0033	&	1953.170	&	1953--1955 (10 eclipse plates)	\\
$T_{5}$	&	2443491.7174	$\pm$	0.0020	&	1977.964	&	Robinson et al. (1982)	\\
$T_{6}$	&	2443492.7150	$\pm$	0.0006	&	1977.966	&	Robinson et al. (1982)	\\
$T_{7}$	&	2443518.7546	$\pm$	0.0002	&	1978.038	&	Robinson et al. (1982)	\\
$T_{8}$	&	2444551.9103	$\pm$	0.0002	&	1980.866	&	Robinson et al. (1982)	\\
$T_{9}$	&	2444552.9104	$\pm$	0.0002	&	1980.869	&	Robinson et al. (1982)	\\
$T_{10}$	&	2444553.9131	$\pm$	0.0002	&	1980.872	&	Robinson et al. (1982)	\\
$T_{11}$	&	2444555.9140	$\pm$	0.0002	&	1980.877	&	Robinson et al. (1982)	\\
$T_{12}$	&	2444637.6977	$\pm$	0.0002	&	1981.101	&	Robinson et al. (1982)	\\
$T_{13}$	&	2445323.6875	$\pm$	0.0002	&	1982.979	&	Seitter (1984)	\\
$T_{14}$	&	2449740.3758	$\pm$	0.0002	&	1995.072	&	Smith et al. (1998)	\\
$T_{15}$	&	2454891.4477	$\pm$	0.0003	&	2009.174	&	AAVSO (BDG)	\\
$T_{16}$	&	2454892.4488	$\pm$	0.0002	&	2009.177	&	AAVSO (BDG)	\\
$T_{17}$	&	2455238.2786	$\pm$	0.0003	&	2010.124	&	AAVSO (BDG)	\\
$T_{18}$	&	2455257.3060	$\pm$	0.0002	&	2010.176	&	AAVSO (BDG)	\\
$T_{19}$	&	2455260.3107	$\pm$	0.0002	&	2010.184	&	AAVSO (BDG)	\\
$T_{20}$	&	2455277.3361	$\pm$	0.0002	&	2010.231	&	AAVSO (BDG)	\\
$T_{21}$	&	2455968.3302	$\pm$	0.0002	&	2012.123	&	AAVSO (BDG)	\\
$T_{22}$	&	2455987.3564	$\pm$	0.0003	&	2012.175	&	AAVSO (BDG)	\\
$T_{23}$	&	2456001.3776	$\pm$	0.0004	&	2012.213	&	AAVSO (BDG)	\\
$T_{24}$	&	2456011.3915	$\pm$	0.0003	&	2012.241	&	AAVSO (BDG)	\\
$T_{25}$	&	2456294.4653	$\pm$	0.0003	&	2013.016	&	AAVSO (BDG)	\\
$T_{26}$	&	2456338.5279	$\pm$	0.0002	&	2013.136	&	AAVSO (BDG)	\\
$T_{27}$	&	2456661.6609	$\pm$	0.0002	&	2014.021	&	Steward Obs. 	\\
$T_{28}$	&	2456684.3589	$\pm$	0.0002	&	2014.083	&	AAVSO (BDG)	\\
$T_{29}$	&	2456707.3913	$\pm$	0.0002	&	2014.146	&	AAVSO (BDG)	\\
$T_{30}$	&	2456725.4167	$\pm$	0.0003	&	2014.196	&	AAVSO (BDG)	\\
$T_{31}$	&	2457011.4946	$\pm$	0.0002	&	2014.979	&	AAVSO (BDG)	\\
$T_{32}$	&	2457017.5036	$\pm$	0.0002	&	2014.995	&	AAVSO (BDG)	\\
$T_{33}$	&	2457020.5084	$\pm$	0.0002	&	2015.003	&	AAVSO (BDG)	\\
$T_{34}$	&	2457395.3798	$\pm$	0.0002	&	2016.030	&	AAVSO (BDG)	\\
$T_{35}$	&	2457395.3798	$\pm$	0.0002	&	2016.030	&	AAVSO (BDG)	\\
$T_{36}$	&	2457407.3977	$\pm$	0.0002	&	2016.063	&	AAVSO (BDG)	\\
$T_{37}$	&	2457798.2939	$\pm$	0.0004	&	2017.133	&	AAVSO (BDG)	\\
$T_{38}$	&	2457803.2996	$\pm$	0.0002	&	2017.147	&	AAVSO (BDG)	\\
$T_{39}$	&	2457815.3171	$\pm$	0.0002	&	2017.180	&	AAVSO (BDG)	\\
$T_{40}$	&	2457827.3351	$\pm$	0.0001	&	2017.212	&	AAVSO (BDG)	\\
$T_{41}$	&	2457828.3347	$\pm$	0.0002	&	2017.215	&	AAVSO (BDG)	\\
$T_{42}$	&	2458125.4275	$\pm$	0.0003	&	2018.029	&	AAVSO (BDG)	\\
$T_{43}$	&	2458137.4467	$\pm$	0.0003	&	2018.061	&	AAVSO (BDG)	\\
$T_{44}$	&	2458151.4659	$\pm$	0.0002	&	2018.100	&	AAVSO (BDG)	\\
$T_{45}$	&	2458161.4803	$\pm$	0.0002	&	2018.127	&	AAVSO (BDG)	\\
		\hline
	\end{tabular}
\end{table}

BT Mon has deep eclipses (Robinson et al. 1982), and these are perfect for accurate timings of the conjunction for the binary orbit.  I have collected many eclipse times from the literature, and these are displayed in Table 3 (with the same format as Table 1).  This includes 8 post-nova times from Robinson et al. (1982), the composite eclipse time from Seitter (1984), and the one time reported by Smith, Dhillon \& Marsh (1998).  For pre-nova eclipse times, Wachmann (1968) reports on four plates, all from immediately before the nova eruption in 1939, where the star is greatly fainter than maximum and is in eclipse.

To bring the $O-C$ curve forward from 1995, I have derived a minimum time from the time series made with the Steward 61-inch telescope in 2014.  Further, I have extracted the many time series from 2009--2018 observed by David Boyd (AAVSO observer code `BDG') from Wantage in England.  Boyd's light curves have 2556 magnitudes with an unfiltered CCD, covering 31 eclipses.

Another source of post-eruption eclipse times comes from the tail of the eruption light curve.  These magnitudes are from 46 Harvard plates 1939--1951 and 263 plates 1939--1962 from Wachmann (1968).  Eclipses will start becoming visible in the tail of the eruption light curve after the nova shell becomes mostly transparent and dims enough so that the central binary dominates the light.  So we have to de-trend the eruption light curve, and the eclipses stand out as plates well below this trend.  These eclipses had already been identified in SP83.  The extraction of eclipse times comes from fitting the de-trended light curve for various intervals to an eclipsing binary profile.  No eclipses are visible before 1941 (when the nova shell dominates), and there are too few plates from 1942--1947 and 1956--1962 for an eclipse to be confidently measured,  In 1941, I have 27 plates including 5 showing eclipses, in 1948 I have 9 plates with one showing an eclipse, in 1950--1951 I have 23 plates of which 2 show eclipses, and in 1953--1955 I have 128 plates with ten showing eclipses.  Critically, the light curves for these four intervals all show gaps in the phased light curve over the duration of the eclipse, with this serving to define and confirm the eclipse phase.  The derived eclipse times (see Table 3) have good utility to define the $\dot{P}$ and to take the observed $O-C$ curve back close to the nova so that little extrapolation is needed.

The last source of eclipse times is six plates from Harvard and the Maria Mitchell observatories that show BT Mon in eclipse.

In all, I have 45 post-eruption eclipse times, well-spaced from 1941--2018, plus 10 pre-eruption times from single archival plates showing BT Mon in eclipse.

\subsection{BT Mon Post-Eruption Orbital Period}

The post-eruption eclipse times can be put into an $O-C$ diagram, and for this I use the fiducial ephemeris of Robinson et al. (1982) with a period of 0.3338141 days and an epoch of HJD 2443491.7155 (see Fig. 5).  We see a simple situation, with small scatter and a blatant parabolic term.  The parabolic term is forced whether we just consider the interval 1941--1995 or the interval 1977--2018, while the set of timings from Boyd 2009--2018 also significantly shows the curvature.  The $O-C$ curve looks to be consistent with a simple parabola (i.e., a constant $\dot{P}$), with little in the way of short-term effects.

The chi-square fit gives a period of 0.33381389	$\pm$0.00000003 days at an epoch of HJD 2443491.7159$\pm$0.0004.  The best fit $\dot{P}$ is -(2.10$\pm$0.14)$\times$10$^{-11}$ days/cycle, with the negative sign showing that the orbital period in quiescence is decreasing over the years.  Extrapolating back to the time of eruption, we have $E_0$=2429516.2452	$\pm$0.0018 and $P_{post}$=0.33381477$\pm$0.00000007 days.

\begin{figure}
	\includegraphics[width=\columnwidth]{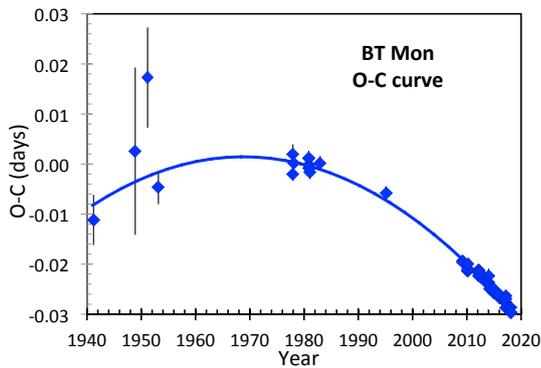}
    \caption{$O-C$ curve for post-eruption eclipse times for BT Mon.  The eclipse times include 4 from plates in the tail of the eruption (1941--1955), 7 from Robinson, Nather, \& Kepler (1982), 1 composite time from Seitter (1984), 1 time from Smith et al. (1998), 31 from time series by David Boyd, and 1 from the Steward Observatory 61-inch.  This $O-C$ curve is based on the fiducial ephemeris in Robinson, Nather, \& Kepler.  What we see is a nice parabola with little noise.  The $\dot{P}$ value is negative, with the steady period decreasing, as expected for magnetic braking.  We see that the extrapolation back to 1939.689 can be done confidently and accurately for both $P_{post}$ and $E_0$.}
\end{figure}

\subsection{BT Mon Pre-Eruption Magnitudes}

\begin{table}
	\centering
	\caption{BT Mon pre-eruption B magnitudes}
	\begin{tabular}{llll} 
		\hline
		Mid-exposure (HJD)	&   Year   & B (mag) & Source  \\
		\hline

2416938.5401	&	1905.253	&	15.76	&	HCO (I 32966)	\\
2419370.8329	&	1911.908	&	15.84	&	HCO (MC 1451)	\\
2419391.7594	&	1911.966	&	15.46	&	HCO (MC 1485)	\\
2420435.9047	&	1914.824	&	16.31	&	HCO (MC 6746)	\\
2420451.8595	&	1914.868	&	16.45	&	HCO (MC 6920)	\\
2420808.7937	&	1915.849	&	15.45	&	HCO (MC 9588)	\\
2421545.8470	&	1917.867	&	15.55	&	HCO (MC 14291)	\\
2423062.7012	&	1922.020	&	15.49	&	HCO (MC 18327)	\\
2423470.5948	&	1923.133	&	15.55	&	HCO (MC 19555)	\\
2423499.5285	&	1923.213	&	15.42	&	HCO (MC 19636)	\\
2424084.8811	&	1924.815	&	15.69	&	HCO (MC 21086)	\\
2424124.8072	&	1924.924	&	15.53	&	HCO (MC 21171)	\\
2424857.5254	&	1926.934	&	15.49	&	HCO (MF 10995)	\\
2424912.6796	&	1927.084	&	15.51	&	HCO (MC 22357)	\\
2425176.9062	&	1927.808	&	15.77	&	HCO (MC 22810)	\\
2425239.2759	&	1927.979	&	15.61	&	HCO (MF 11480)	\\
2425239.7635	&	1927.980	&	15.60	&	HCO (MC 23017)	\\
2425327.2758	&	1928.220	&	15.50	&	HCO (MF 11456)	\\
2425587.8323	&	1928.933	&	15.60	&	HCO (MC 23875)	\\
2425595.7777	&	1928.955	&	15.66	&	HCO (MA 2359)	\\
2425595.8345	&	1928.955	&	15.54	&	HCO (MA 2360)	\\
2425596.5274	&	1928.957	&	15.45	&	HCO (MF 12669)	\\
2425615.4916	&	1929.009	&	15.59	&	HCO (MF 12708)	\\
2425644.4138	&	1929.088	&	15.38	&	HCO (MF 12776)	\\
2425943.5590	&	1929.907	&	$>$16.8	&	HCO (MF 13775)	\\
2425972.5204	&	1929.986	&	15.55	&	HCO (MF 13863)	\\
2426011.3421	&	1930.093	&	16.03	&	HCO (MF 13924)	\\
2426035.3426	&	1930.158	&	16.78	&	HCO (MF 13964)	\\
2426065.3035	&	1930.240	&	15.28	&	HCO (MF 14077)	\\
2426663.5517	&	1931.878	&	16.00	&	HCO (B 55622)	\\
2426710.5116	&	1932.007	&	15.66	&	HCO (MF 16327)	\\
2426770.3500	&	1932.171	&	15.50	&	HCO (MF 16494)	\\
2426772.5350	&	1932.177	&	15.45	&	HCO (MC 25918)	\\
2426794.2741	&	1932.236	&	15.48	&	HCO (MF 16553)	\\
2426831.2426	&	1932.337	&	$>$16.1	&	HCO (MF 16721)	\\
2427070.3618	&	1932.992	&	15.42	&	HCO (MF 17677)	\\
2427095.3099	&	1933.060	&	15.81	&	HCO (MF 17735)	\\
2427397.8369	&	1933.8887	&	15.59	&	MMO (NA1601)	\\
2427397.8899	&	1933.8888	&	15.48	&	MMO (NA1602)	\\
2427397.9279	&	1933.8889	&	$>$16.45	&	MMO (NA1603)	\\
2427417.6317	&	1933.9429	&	$>$15.82	&	MMO (NA1608)	\\
2427417.7682	&	1933.9432	&	15.56	&	MMO (NA1611)	\\
2427425.5951	&	1933.9647	&	15.91	&	MMO (NA1614)	\\
2427425.6481	&	1933.9648	&	16.19	&	MMO (NA1615)	\\
2427425.6971	&	1933.9649	&	16.35	&	MMO (NA1616)	\\
2427425.7411	&	1933.9651	&	$>$15.63	&	MMO (NA1617)	\\
2427425.7861	&	1933.9652	&	$>$15.63	&	MMO (NA1618)	\\
2427425.8311	&	1933.9653	&	$>$15.56	&	MMO (NA1619)	\\
2427425.8762	&	1933.9654	&	$>$14.34	&	MMO (NA1620)	\\
2427456.4126	&	1934.049	&	15.66	&	HCO (RB 4807)	\\
2427473.6217	&	1934.096	&	15.57	&	HCO (RH 5722)	\\
2427478.6119	&	1934.110	&	15.37	&	HCO (RH 5757)	\\
2427504.3609	&	1934.180	&	15.45	&	HCO (MF 18992)	\\
2427812.4360	&	1935.024	&	15.36	&	HCO (RB 5877)	\\
2427840.6521	&	1935.101	&	15.47	&	HCO (RH 6443)	\\
2428100.8801	&	1935.813	&	15.63	&	HCO (MC 27944)	\\
2428110.8647	&	1935.841	&	16.12	&	HCO (RH 6887)	\\
2428161.4197	&	1935.979	&	15.83	&	HCO (B 60450)	\\
2428250.2662	&	1936.222	&	15.50	&	HCO (RB 6689)	\\
2428582.7268	&	1937.133	&	15.22	&	HCO (MA 6145)	\\
2428601.3496	&	1937.184	&	15.68	&	HCO (B 61709)	\\
2428633.2553	&	1937.271	&	16.01	&	HCO (B 61822)	\\
2428838.8612	&	1937.834	&	15.37	&	HCO (RH 7836)	\\
		\hline
	\end{tabular}
\end{table}

\begin{table}
	\centering
	\contcaption{Pre-eruption light curve for BT Mon}
	\label{tab:continued}
	\begin{tabular}{lllll} 
		\hline
		Mid-exposure (HJD)	&   Year   & B (mag) & Source  \\
		\hline
2428904.5563	&	1938.014	&	15.15	&	HCO (RB 7690)	\\
2428905.3468	&	1938.016	&	15.53	&	HCO (B 62730)	\\
2428919.6680	&	1938.055	&	15.19	&	HCO (RH 8046)	\\
2428926.6504	&	1938.074	&	15.28	&	HCO (RH 8058)	\\
2428927.5988	&	1938.077	&	15.38	&	HCO (RH 8063)	\\
2428927.6418	&	1938.077	&	15.32	&	HCO (RH 8064)	\\
2428951.5513	&	1938.143	&	15.48	&	HCO (RH 8120)	\\
2428959.3655	&	1938.164	&	15.42	&	HCO (RB 7738)	\\
2428985.5471	&	1938.236	&	15.58	&	HCO (RH 8210)	\\
2429204.5221	&	1938.835	&	15.53	&	HCO (B 63814)	\\
2429219.5373	&	1938.876	&	15.40	&	HCO (RB 8369)	\\
2429224.5399	&	1938.890	&	15.88	&	HCO (RB 8389)	\\
2429230.7825	&	1938.907	&	15.34	&	HCO (RH 8614)	\\
2429250.4950	&	1938.961	&	15.52	&	Wachmann (1968)	\\
2429251.5851	&	1938.964	&	15.52	&	Wachmann (1968)	\\
2429290.6324	&	1939.071	&	15.32	&	HCO (RH 8743)	\\
2429302.3642	&	1939.103	&	15.52	&	Wachmann (1968)	\\
2429302.3842	&	1939.103	&	15.69	&	Wachmann (1968)	\\
2429308.3838	&	1939.119	&	15.52	&	Wachmann (1968)	\\
2429318.5535	&	1939.147	&	15.38	&	HCO (BM 1199)	\\
2429334.3519	&	1939.191	&	15.69	&	Wachmann (1968)	\\
2429335.3533	&	1939.193	&	15.46	&	HCO (RB 8504)	\\
2429340.3113	&	1939.207	&	16.35	&	Wachmann (1968)	\\
2429344.3210	&	1939.218	&	16.57	&	Wachmann (1968)	\\
2429364.3492	&	1939.273	&	$>$16.38	&	Wachmann (1968)	\\
2429365.2608	&	1939.275	&	15.48	&	HCO (B 64118)	\\
2429365.3491	&	1939.275	&	$>$16.38	&	Wachmann (1968)	\\
		\hline
	\end{tabular}
\end{table}

Wachmann (1968) reported on 280 magnitudes of BT Mon from 1938--1962 taken from plates with the  Lippert-Astrograph at Hamburg Observatory.  Many of the magnitudes have already been used to find eclipses in the tail of the nova light curve.  Ten pre-eruption observations are reported, and these include four plates that show BT Mon greatly fainter than normal, i.e., in eclipse.  Wachmann identifies his comparison stars and tells us his adopted magnitudes, and from this, we can transform his quoted magnitudes into modern B magnitudes (see Johnson et al. 2014).  

The Harvard archive has many plates showing BT Mon before its 1939 nova eruption, and I have measured the magnitudes from these plates with up to five independent times.  I now have a total of 68 useful plates, all with magnitudes placed into the Johnson B system.  (My measures reported in SP83 were in the magnitude system of Wachmann's sequence, and these have been transformed to B in the same way as for Wachmann's data.)  I have not included the plates that I have measured only once, and have not seen since 1983.  I have excluded the plate RH 8190, for which I now see a small scratch on the target image that would affect the photometry.  I have made heliocentric corrections for the times of mid-exposure, and I have used the times of mid-exposures from the DASCH reconstructions from the logbooks.

The Maria Mitchell Observatory (MMO) has a series of roughly 8000 plates (taken with the 7.5-inch Alvan Clark telescope) covering much of the sky from 1913--1995 as viewed from Nantucket Island in Massachussetts.  The plates have been digitized, but direct visual examination of the plates goes about one magnitude deeper, and this is required to catch BT Mon in quiescence.  The glass plates are now stored at the  {\it Pisgah Astronomical Research Institute} in North Carolina.  I found 12 useful plates taken as times series on three nights in November and December of 1933.  The exposure times range from 30--75 minutes, with seven being close to 60 minutes.  I estimated the magnitudes of BT Mon by comparison with nearby stars whose B is known from APASS.  The first and third nights record BT Mon in eclipse.

For the last night of MMO plates, with an apparent eclipse recorded, a substantial problem arises because neither the plate logs nor the plate jackets record the times in any way.  Fortunately, the exposure times for the seven consecutive BT Mon exposures (plus 3 minutes for each plate change) fits exactly into the available time that night at the target rises from 22$\degr$ above the horizon in the east to pass the meridian and then sets down to 22$\degr$ above the horizon in the west.  Given this forced timeline, the HJD of the middle of each exposure can be estimated to an accuracy of roughly $\pm$0.006 days.

In all, I have 90 pre-eruption B magnitudes for BT Mon, from 1905 to just a few months before the eruption.  These are listed in full in Table 4.  I have ten plates that certainly show BT Mon in eclipse, plus several more that show the start and middle of the ingress and egress.  This is many more eclipses than needed to establish a very confident and accurate $P_{pre}$.

\subsection{BT Mon Pre-Eruption Orbital Period}

I have determined $P_{pre}$ from the same three methods as has been done with QZ Aur (Schaefer et al. 2019) and DQ Her (see Section 5.4).

For the periodogram, I calculate the $F$ parameter for all 90 pre-eruption magnitudes for a range of trial periods within 1000 ppm of $P_{post}$.  The parameter is as used in an F-test comparing the two hypotheses that the folded light curve has the correct period and the post-eruption eclipse shape, versus the hypothesis that the period is wrong.  Figure 6 displays this periodogram.  This plot is practically identical to that already given in SP83.  (The new data includes two eclipses that make the peak higher, while the at-maximum points further beat down the noise.  The change in the vertical scale is only because the 1983 work used binned light curves due to limited computational speed in olden times.)  The periodogram shows a prominent, isolated, and high peak at a period of 0.3338016 days.  The height of the peak shows the periodicity to be very significant.  The lack of any other significant peaks, in this exhaustive test over the entire physically possible range, is a proof that there is no possible alternative periodicity.  (The second highest peak is an ordinary 25-year alias between the 1914 and 1939 eclipses.)  A further point from Fig. 6 is that the period from the pre-eruption peak is certainly smaller than the post-eruption period, so the orbital period {\it increased} across the 1939 eruption.  Thus, a glance at the plot from the first method proves that $P_{pre}$ is unique and highly-significant, and that $\Delta$P is positive.

\begin{figure}
	\includegraphics[width=\columnwidth]{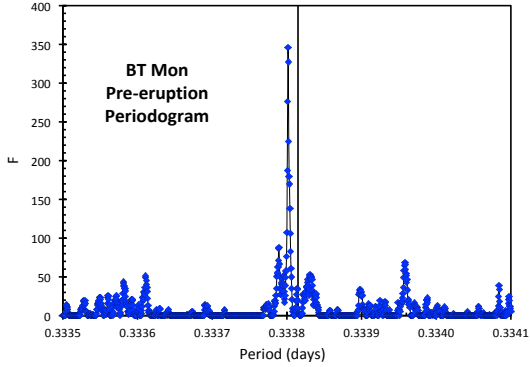}
    \caption{Periodogram for BT Mon before the 1939 eruption.  We see a very-high peak at a period of 0.3338016 days.  The vertical line represents the post-eruption orbital period.  We can quickly see three main points from this plot.  First, the peak is very highly significant, so $P_{pre}$ is the true period.  Second, the plot shows an exhaustive examination of all possible $P_{pre}$ values, so there can be no other alternative $P_{pre}$ somehow missed.  Third, we see that $P_{pre}<P_{post}$, so the period {\it increased} across the 1939 eruption.}
\end{figure}

For the second method, looking for the common divisor of all the times between eclipses, we get the same period.  For adjacent eclipse times in Table 3, we have intervals that are 3, 12, 48, 60, and 83 cycles long, so we can keep the cycle count to 83.  The pre-eruption period calculated for this 83 cycle interval is $(T_{-5}-T_{-6})/\Delta N_{-5,-6}$ or 0.33398$\pm$0.00038 days.  From the first to last of the Wachmann eclipses ($T_{-4}$ to $T_{-1}$), we get 0.33384$\pm$0.00038 days.  For $T_{-8}$ to $T_{-7}$, we get $\Delta N_{-8,-7}$=275 and 0.33376$\pm$0.00008 days.  (This period for 275 cycles is consistent with the period from other intervals, whereas if we had adopted 274 or 276 cycles instead, the derived periods of 0.33498$\pm$0.00005 or 0.33255$\pm$0.00005 days would be rejected as being incompatible with the other period constraints already calculated.  This is the basic logic that allows us to keep the cycle count as we push to longer and longer intervals.)  These are independent measures, and a weighted average is 0.333772$\pm$0.000077 days.  This is adequately accurate to get the correct cycle count for longer time intervals.  For the interval from $T_{-8}$ to $T_{-6}$, $\Delta N_{-8,-6}$ can only be 4357, and the period is 0.3338005$\pm$0.0000033 days.  We can check and improve this with $\Delta N_{-6,-3}$=5831 and period 0.3338009$\pm$0.0000024 days.  Extending to a larger interval, where $\Delta N_{-7,-4}$ can only be 9901 so as to be consistent with the above periods, we get a period of 0.3338015$\pm$0.0000029 days.  Doubling the interval again for $\Delta N_{-9,-6}$=20809, we have a period of 0.3338012$\pm$0.0000015 days.  The longest time interval of 26751 cycles gives the best period of  $(T_{-10}-T_{-1})/\Delta N_{-10,-1}$ or 0.3337985$\pm$0.0000014 days.  So just simple arithmetic on the ten pre-eruption eclipse times returns a robust and confident and accurate period.

For the third method, the chi-square fitting to the modern eclipse profile, I have performed the fit on the   90 pre-eruption magnitudes.  The eclipse profile has a flat maximum at B=15.55, and the deepest eclipse reaching to B=16.6, the HWZI of the eclipse is 0.09 in phase, and a HWHM of 0.06 in phase.  The one-sigma for the photometry is 0.19 mag, as determined by the RMS scatter of the at-maximum magnitudes from phase 0.10--0.90.  There is only a soft constraint on $\dot{P}$ (because the eclipse times from single plates have greatly lower accuracy than the post-eruption eclipse timings), and I get (-8$\pm$3)$\times$10$^{-11}$ days/cycle.  The best fit period is 0.33380144$\pm$0.00000025 days at the time of the epoch 2426844.1590$\pm$0.0015.  When extrapolated forward to the time of the nova, the eclipse epoch is $E_0$=2429516.2370$\pm$0.0016 and $P_{pre}$=0.3338008$\pm$0.0000004 days.

\begin{figure}
	\includegraphics[width=\columnwidth]{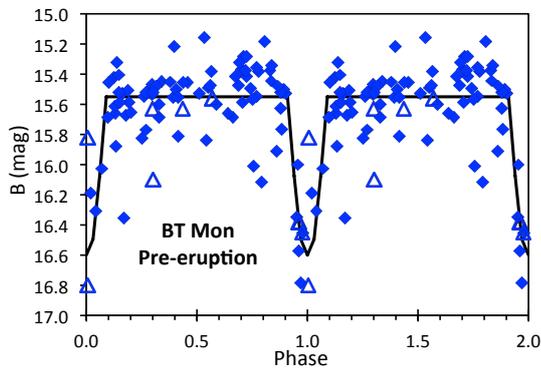}
    \caption{Folded light curve for BT Mon before the nova eruption.  Here are 90 pre-eruption magnitudes folded on the best fit pre-eruption orbital period.  The upper limits are represented by triangles, and all the magnitudes are plotted twice, once for phase 0.0--1.0 and again for 1.0--2.0.  We see zero at-maximum plates (B$<$16.0) between phases -0.09 and +0.09 (also from 0.91 to 1.09), and we see all ten eclipse plates (with B$>$16.2) between phases -0.05 to +0.05 (also 0.95 to 1.05).}
\end{figure}

The best fit folded light curve for the 90 pre-eruption magnitudes of BT Mon are shown in Fig. 7, along with the eclipse profile used as the model.  We see a classic eclipsing binary light curve.  Critically, we see zero plates brighter than B=16.0 over the eclipse phase range of -0.09 to +0.09, and we see all ten eclipse plates (with B$>$16.2) in the phase range -0.05 to +0.05.  And we see two plates with B close to 16.0 with phases near -0.06 and +0.06, simply showing the ingress and egress.  This provides strong and quick visible proof that the $P_{pre}$ is correct.

Three points in Fig. 7 are from plates with B fainter than 16.0 for phases well away from any eclipse.  Plates RH 6887 and MF 16721 were both measured four times, while plate B 61822 was measured twice, all with closely consistent measures, and with my notes pointing out (for some of these measures) that the faintness was recognized and confirmed.  So measurement error is unlikely, while there is no precedent (out of many long times series in modern times) for BT Mon away-from-eclipses being faint by 0.4 mag or more.  It is always possible that the plate times were mis-recorded on both the logbooks and plate jackets, but such is unlikely.  However, all three of these points are just 2--3 sigma from the out-of-eclipse mean.  So the easiest explanation is that we expect $\sim$5\% of the 90 plates to deviate by more than 2-sigma from the model, and these plates are simply that tail of the distribution.

One anomaly is that MMO plate NA 1616 has B=16.35 (i.e., apparently in eclipse) at a phase far from any eclipse epoch.  Now, the MMO plates from that night only have times determined by barely fitting the exposures into the apparition of BT Mon transiting the sky, so there is some real uncertainty in the time.  An easy explanation would be to have plate NA 1616 exposed 0.05 days earlier, but this pushes the first exposure of BT Mon (plate NA 1614) to much too low an altitude.  In any case, such a shift in time would also shift plate NA 1615, and this plate also shows BT Mon in eclipse.  But they both cannot show BT Mon faint, because the two back-to-back plates have their middles separated by 71 minutes, while the HWHM of the eclipse profile is 29 minutes.  So something is apparently wrong with plate NA 1616.  (This is exactly the same problem as with the pair of Vatican plates for QZ Aur, see Schaefer et al. 2019.  But in that case, the telescope was a double-astrograph designed for taking two simultaneous exposures, so a simple mis-labelling of time onto the plate jacket is the easy and likely solution.)  Perhaps the easiest solution for plate NA 1616 is simply that I made an error and estimated the magnitude much fainter than depicted on the plate.  This is an unlikely solution, but I did measure the plate only once.  The solution to the NA1616 anomaly is that this plate has some sort of a measurement error, and such happens often enough in practice.  Indeed, in dozens of studies of constant stars and out-of-eclipse binaries, the $>$4-sigma rate of outliers varies from 0.01--1.0 per cent for Harvard plates.  (Even after ordinary quality checks, the ASAS CCD light curves have $>$4-sigma outlier rates from 0.1--5 percent for uncrowded stars, while the AAVSO visual light curves have $>$4-sigma outlier rates of 0.05--1.0 per cent.  That is, photographic photometry has comparable outlier rates as does CCD and visual photometry.)  So the one anomaly of NA 1616 is not worrisome, as such outliers are ordinary and common enough.  In the meantime, the use of the NA 1616 magnitude does not change the chi-square fit because it is on the flat part of the model light curve, NA 1616 is included in the periodogram with no new peaks coming up in the noise, and the use of the NA 1616 magnitude does not significantly change the eclipse time analysis.  That is, the existence of the anomalous NA 1616 magnitude is not worrisome, and does not change the derived $P_{pre}$ in any of the three analyses.

The $E_0$ from the pre-eruption data is just 0.0082 days different from the $E_0$ derived from the completely independent post-eruption data.  This is just 1/40 of the period.  The chances are small for this happening if $P_{pre}$ is spurious.

So now we have three greatly-different methods of pulling out the pre-eruption orbital period, and they are all consistent.  The most accurate measure (as it uses all the information plus a $\dot{P}$) is from the chi-square fit, so $P_{pre}$=0.3338008$\pm$0.0000004 days.

\subsection{BT Mon Joint Fit}

The best value for $\Delta$P and $\dot{P}$ (plus their errors) comes from the simultaneous chi-square fit of the 90 pre-eruption magnitudes and the 45 post-eruption eclipse times.  The $E_0$ and $\dot{P}$ values for the pre- and post-eruptions fits are similar, so the joint fit will not be forcing any substantial changes.  For the joint fit, I presume that $\dot{P}$ remains constant throughout the time since the 1800s.

The joint best fit has $E_0$=2429516.2413$\pm$0.0018 in HJD and $\dot{P}$=(-2.3$\pm$0.1)$\times$10$^{-11}$ days/cycle.  The negative $\dot{P}$ shows that the orbital period has a steady {\it decrease} over the decades.  The best fit $P_{pre}$=0.33380167$\pm$0.00000011 days just before the 1939 eruption.  The best fit $P_{post}$=0.33381490$\pm$0.00000006 days just after the 1939 eruption.

\subsection{$\Delta$P for BT Mon}

The period change has been very confidently measured.  With the periods from the joint fit, the period change is $\Delta$P=+0.00001323$\pm$0.00000017 days.  The positive sign shows that the orbital period of BT Mon has {\it increased} across the 1939 eruption.  The fractional change is accurately measured as $\Delta$P/P=39.6$\pm$0.5 ppm.  This is the same result as in SP83, yet with an order-of-magnitude smaller error bar.

\section{Testing Models}

This paper presents new measures of $\Delta$P and $\dot{P}$ for CNe DQ Her and BT Mon.  The measured $P_{pre}$ are robust.  These values are summarized in Table 5.  These measures can be compared to the predictions of mechanisms for $\Delta$P and then for the two venerable models for CV evolution.

\subsection{$\Delta$P Mechanisms}

The usual mass loss mechanism always produces positive $\Delta$P (for $q$$>$1 or so).  FAML will always produce a negative $\Delta$P.  For CNe, the sum of these two mechanisms must always be positive.  (See Table 5 for a breakdown on $\Delta$P$_{ml}$/P and $\Delta$P$_{FAML}$/P.)  The magnetic braking will always produce a negative $\Delta$P.  The sum of all three effects will always be positive, except in the case where the companion star has a very high magnetic field (Martin et al. 2011).  So, except for the unlikely case of a very high magnetic field, the three usual effects will have $\Delta$P substantially greater than zero.

This raises a problem since DQ Her has a negative value.  This proves that there must be some additional mechanism operating to decrease $P$ substantially across a nova eruption.  

The only candidate mechanism is that recently proposed by J. Frank (Schaefer et al. 2019), where the nova ejects an asymmetric shell, which provides either a positive or negative kick to the white dwarf, making for a sudden change in $P$.  For the system parameters, I have calculated the maximum value of $\Delta$P$_{ml}$/P, i.e., for $\xi$=$\pm$1 (see Table 5).  We see that it is very easy to get the DQ Her value with a modest negative $\xi$.  That is, there must be some additional mechanism making the sharp drop in $P$ in 1934, and this could well be the Frank mechanism of asymmetric shell ejections.

\subsection{Hibernation Model}

The Hibernation model is driven-by and requires that $\Delta$P$>$0.  DQ Her certainly has a robust measure of $\Delta$P$<$0.  So DQ Her provides a confident counterexample for Hibernation.

BT Mon has a positive $\Delta$P.  But it is not large enough to drive anything that we would care to call as hibernation.  That is, a small positive $\Delta$P will indeed make the orbit expand slightly, and this should make the accretion rate go down somewhat.  But a small drop in brightness (from before the eruption until long after the eruption), would not be called hibernation.  To be called `hibernation', the accretion must largely turn off.  The criterion for a turn-off certainly corresponds to an absolute magnitude fainter than the lowest luminosity for a system recognized as a CV, i.e., $M_V$$>$12 mag.  If we wanted to have a mild-Hibernation case, we would not call it a significant enough unless there is a drop in $M_V$ by at least 5 mag.  For the BT Mon system parameters, Table 5 gives the minimum $\Delta$P/P required to match these two criteria.

For BT Mon to go into hibernation, the system must have $\Delta$P/P$>$ 2500 ppm.  Even for a stunted hibernation case, the system would have to have $\Delta$P/P$>$1580 ppm.  But the observed $\Delta$P/P is 40$\times$ too small for even the stunted hibernation case.  Thus, BT Mon certainly is not going into hibernation.  This is a robust demonstration that the Hibernation model is not working for this one system.

V1017 Sgr has $\Delta$P/P = -273$\pm$61 ppm (Salazar et al. 2017).  QZ Aur has has $\Delta$P/P = -290.71$\pm$0.28 ppm (Schaefer et al. 2019).  With both having negative values, the Hibernation model has two additional confident counterexamples.

So we have four-out-of-four CNe measured where Hibernation is certainly not working.  DQ Her, BT Mon, V1017 Sgr, and QZ Aur have properties that span the usual range for CNe, and being consistent with a random sampling from CNe.  We could even call DQ Her as the prototype CN.  So it is not like we have some biased CN sample that is somehow selected against Hibernation.  Rather, with four-out-of-four ordinary CNe refuting Hibernation, we know that Hibernation must be uncommon amongst CNe.  With Hibernation being claimed to solve demographic questions, we know that Hibernation has failed. 

\subsection{Magnetic Braking Model}

\begin{table}
	\centering
	\caption{Observed Period Changes Versus Predictions}
	\begin{tabular}{lll} 
		\hline
	&	DQ Her			&	BT Mon			\\
		\hline
\underline{{\bf $\Delta$P Program Results:}}	&				&				\\
~~Pre-eruption plates	&	52 (3 eclipses)			&	90 (10 eclipses)			\\
~~$P_{pre}$ (days)	&	0.1936217610			&	0.33380167			\\
	&		$\pm$	0.0000000055	&		$\pm$	0.00000011	\\
~~$P_{post}$ (days)	&	0.1936208977			&	0.33381490			\\
	&		$\pm$	0.0000000017	&		$\pm$	0.00000006	\\
~~$\Delta$P (days)	&	-0.0000008633			&	0.00001323			\\
	&		$\pm$	0.0000000058	&		$\pm$	0.00000013	\\
~~$\Delta$P/P (ppm)	&	-4.46	$\pm$	0.03	&	$+$39.6	$\pm$	0.5	\\
~~$\dot{P}$ ($10^{-11}$ days/cycle)	&	0.00	$\pm$	0.02	&	-2.3	$\pm$	0.1	\\
\underline{{\bf $\Delta$P calculations:}}	&				&				\\
~~$\Delta$P$_{ml}$/P (ppm)	&	200			&	21			\\
~~$\Delta$P$_{FAML}$/P (ppm)	&	-5.9			&	-0.2			\\
~~Max. $\Delta$P$_{jet}$/P (ppm)	&	$\pm$560			&	$\pm$160			\\
~~$\Delta$P$_{5mag}$ (ppm)	&	1030			&	1580			\\
~~$\Delta$P$_{M_q=+12}$/P (ppm)	&	1360			&	2500			\\
\underline{{\bf $\dot{P}$ calculations:}}	&				&				\\
~~$\dot{P}_{model}$ ($10^{-11}$ days/cycle)	&	-0.027			&	-0.33			\\
~~$\dot{P}_{mt}$ ($10^{-11}$ days/cycle)	&	0.059			&	0.16			\\
~~$\dot{P}_{mb}$ ($10^{-11}$ days/cycle)	&	-0.078			&	-0.51			\\
~~$\dot{P}_{\Delta P}$ ($10^{-11}$ days/cycle)	&	-0.001			&	0.60			\\

		\hline
	\end{tabular}
\end{table}

The MBM predicts that all CNe will follow a single unique path for $\dot{P}$ as a function of $P$.  In Table 5, $\dot{P}_{model}$ is taken from the best MBM model of Knigge et al. (2011).  This includes the effects of both magnetic braking ($\dot{P}_{mb}$) and of steady mass transfer ($\dot{P}_{mt}$).  The tabulated value of $\dot{P}_{mt}$ is for the observed value of $\dot{M}$, as calculated from equation (2), as opposed to the value based on the MBM model value for $\dot{M}$.  

The measured $\dot{P}$ for DQ Her is consistent with the MBM prediction.  But the measured $\dot{P}$ for BT Mon is greatly different from the MBM prediction.  Further, the measured $\dot{P}$ for QZ Aur ($-$2.84$\pm$0.22 $\times$ 10$^{-11}$ days/cycle, see Schaefer et al. 2019) is greatly different from the MBM prediction ($-$0.54 $\times$ 10$^{-11}$ days/cycle).  (The MBM does extend to systems with long orbital periods, such as for V1017 Sgr.)  This presents a deep problem for the MBM, as its core prediction of the speed of the period decrease is contradicted in two-out-of-three cases.  

The MBM predictions of $\dot{P}$ have never been tested before.  And the deviations are large and highly significant.  With these greatly different observed values of $\dot{P}$, all of the CV demographics would have to change substantially.  This is a serious challenge to MBM.

A second and completely-independent serious challenge to MBM comes because no one has previously considered the effects of the many $\Delta$P shifts for each eruption on the overall evolution.  For the case of BT Mon, we have $\Delta$P of 0.00001323 d that will occur once per eruption, for which I calculated that $\tau_{rec}$ is 2000 years (2.2 million orbits), so the long-time average of period change from this mechanism is $\dot{P}_{\Delta P}$=+0.60$\times$10$^{-11}$ days/cycle.  That is, if BT Mon repeatedly has the same $\Delta$P for each eruption every two millennia or so, then this will contribute a significant and substantial period change with $P$ getting {\it longer}.  This goes in the opposite direction as does the MBM prediction, and this effect is twice as large as $\dot{P}_{model}$.  These effects are independent, so they should be added together.  So MBM should be using $\dot{P}_{model}$+$\dot{P}_{\Delta P}$ for use in its calculations of CV evolution.  With this, the very long term evolution of BT Mon would have a {\it positive} total period change.  That is, BT Mon is not having its period being `braked', but rather it is being accelerated.  (To be sure, the magnetic braking mechanism is certainly operating, it is just that other effects are dominating, so the total effect is opposite that predicted by MBM.)  So the second challenge to MBM is for BT Mon, where we see that MBM is not working, at least for this one eruption cycle.

Further, for the case of QZ Aur, both the observed $\dot{P}$ ($-$2.84$\pm$0.22 all in units of 10$^{-11}$ days/cycle) and $\dot{P}_{model}$ ($-$0.54) are dwarfed by $\dot{P}_{\Delta P}$ ($-$33.9) for $\tau_{rec}$ equal to 300 years.  In this case, the magnetic braking effect is just small noise in the long term period change of QZ Aur.  So again, we have the second strong challenge, where MBM is irrelevant in the face of large $\Delta$P. 

My $\Delta$P program has the first measures for either $\Delta$P or the parabolic term $\dot{P}$ for any CNe.  Now, I have two published CNe (V1017 Sgr and QZ Aur), this paper reports on two more CNe (DQ Her and BT Mon), while the companion paper reports on two more CNe (RR Pic and HR Del).  These 6 CNe are all the systems for which such measures are possible for even far into the future.  With just the first four discussed in this paper, we already have four-out-of-four CNe as solid counterexamples of the venerable Hibernation model of CV evolution.  Further, we have two strong challenges to the MBM, first, where the measured $\dot{P}$ values are in strong disagreement with the MBM predictions for two-out-of-three CNe, plus the second challenge being that the previously unrecognized effects of the sudden $\Delta$P averaged over the eruption cycle dominate over the MBM $\dot{P}$.

\section*{Acknowledgements}

The AAVSO has provided much that was required for this program, including finder charts, data archiving, and comparison star magnitudes (through the APASS program).  Funding for APASS has been provided by the Robert Martin Ayers Sciences Fund.  We have used the DASCH catalogs, thumbnails, and photometry, with the DASCH project having partial support from NSF grants AST-0407380, AST-0909073, and AST-1313370.  We are grateful for the historical observations and conservation made for the archival plates at Harvard, Sonneberg, and Maria Mitchell by many workers over the last 130 years.  For the particular work in this paper, we are thankful for the hospitality and help from Alison Doane, Josh Grindlay, Peter Kroll, and Mike Castelaz.  Chris Johnson observed the time series showing the BT Mon eclipse in 2014 at Steward Observatory.  David Boyd provided many time series centered on eclipses of BT Mon.  I am also very thankful for the myriad of observers recording many eclipses of DQ Her.



\bsp	
\label{lastpage}
\end{document}